\documentclass[11pt]{article}
\usepackage{amssymb,latexsym, amsfonts}  
\usepackage{epsf}                        

\newcommand{\sect}[1]{\setcounter{equation}{0}\section{#1}}
\renewcommand{\theequation}{\arabic{section}.\arabic{equation}}

\def\be{\begin{equation}}
\def\ee{\end{equation}}
\def\bea{\begin{eqnarray}}
\def\eea{\end{eqnarray}}
\def\nn{\nonumber \\}

\def\hsp#1{\hspace{#1}}

\def\part{\partial}
\def\tfrac#1#2{{\textstyle{#1\over #2}}}
\def\half{\tfrac{1}{2}}
\def\x{\times}

\def\Tr{\mbox{Tr}}
\def\STr{\mbox{STr}}

\def\incl{\mbox{i}}

\def\dA{\dot A}
\def\dX{\dot X}

\def\tb{\tilde b}

\def\cD{{\cal D}}
\def\cP{{\cal P}}

\def\la{\lambda}
\def\al{\alpha}
\def\td{\tilde}
\def\inv{^{-1}}

\def\p{\phi}
\newcommand{\Q}{\ensuremath{{\cal Q}}}
\newcommand{\T}{\ensuremath{{\cal T}}}
\newcommand{\J}{\ensuremath{{\cal J}}}
\newcommand{\M}{\ensuremath{{\cal M}}}
\newcommand{\cO}{\ensuremath{{\cal O}}}
\newcommand{\quarter}{\ensuremath{\frac{1}{4}}}
\newcommand{\del}{\ensuremath{\partial}}

\def\ab{{ab}}
\def\a9{{a9}}
\def\mn{{\mu\nu}}
\def\mnr{{\mu\nu\rho}}
\def\mnrl{{\mu\nu\rho\lambda}}

\def\makeatletter{\catcode`\@=11}
\makeatletter
\def\mathbox#1{\hbox{$\m@th#1$}}%
\def\math@ccstyles#1#2#3#4#5#6#7{{\leavevmode
      \setbox0\mathbox{#6#7}%
      \setbox2\mathbox{#4#5}%
      \dimen@ #3%
      \baselineskip\z@\lineskiplimit#1\lineskip\z@
      \vbox{\ialign{##\crcr
             \hfil \kern #2\box2 \hfil\crcr
             \noalign{\kern\dimen@}%
             \hfil\box0\hfil\crcr}}}}
\def\mathaccstyles{\math@ccstyles\maxdimen}
\def\maththroughstyles{\math@ccstyles{-\maxdimen}}
\def\unity%
 {\maththroughstyles{.45\ht0}\z@\displaystyle {\mathchar"006C}\displaystyle 1}
\title{{\bf Dielectric Fundamental Strings in Matrix String Theory}}

\author{{\bf D. Brecher${}^{a,}$}\thanks{email: dominic.brecher@durham.ac.uk}~~,
{\bf B. Janssen${}^{a,}$}\thanks{email: bert.janssen@durham.ac.uk}~~{\bf and}
{\bf Y. Lozano${}^{b,}$}\thanks{email: yolanda@string1.ciencias.uniovi.es} \\
\\ ${}^a$Centre for Particle Theory, Department of Mathematical Sciences, \\
University of Durham, South Road, Durham DH1 3LE,
United Kingdom.\\
\\ ${}^b$Departamento de F{\'\i}sica, Universidad de Oviedo \\
Avda. Calvo Sotelo 18, 33007 Oviedo, Spain.\\
\\ Preprint DCPT-01/79, FFUOV-01/08 \\
hep-th/0112180}

\date{January 16, 2002}

\begin{document}

\maketitle

\begin{abstract}
Matrix string theory is equivalent to type IIA superstring theory
in the light-cone gauge, together with extra degrees of freedom
representing D-brane states.  It is therefore the appropriate framework
in which to study systems of multiple fundamental strings expanding
into higher-dimensional D-branes.  Starting from Matrix theory in a weakly
curved background, we construct the linear couplings of closed string
fields to type IIA Matrix strings.  As a check, we show that at weak
coupling the resulting action reproduces light-cone gauge string
theory in a weakly curved background.  Further dualities give a
type IIB Matrix string theory and a type IIA theory of Matrix strings
with winding.  We comment on the dielectric effect in each of these
theories, giving some explicit solutions describing fundamental strings
expanding into various D$p$-branes.
\end{abstract}

\newpage

\sect{Introduction}

The idea that a collection of branes can undergo an ``expansion'' into
a single higher-dimensional D-brane under the influence of a
background Ramond-Ramond (R-R) potential was first explored by
Emparan~\cite{emparan:97}.  Although his main concern was with the
description of $N$ fundamental strings expanding into a D$p$-brane, he
also realised that D$(p-2)$-branes could undergo such an expansion.
Emparan's analysis of this effect was entirely at the level of the
abelian theory relevant to the description of the single D$p$-brane.
Switching on a background R-R $(p+2)$-form field strength, he found
solutions of the combined Born-Infeld-Chern-Simons theory with
topology $\mathbb{M}^2 \times S^{p-1}$, where $\mathbb{M}^n$ denotes
an $n$-dimensional Minkowski space, and with $N$ units of dissolved
electric flux, corresponding to the fundamental strings.  Similar
solutions, with topology $\mathbb{M}^{p-1} \times S^2$, and with $N$
units of dissolved magnetic flux, corresponding to D$(p-2)$-branes,
are easily found.

It was some years later that a description from the point of view
of the lower-dimensional D$(p-2)$-branes was provided~\cite{myers}.  This
involved an analysis of certain couplings in the non-abelian
Born-Infeld-Chern-Simons action proposed by Taylor and van
Raamsdonk~\cite{TVR2} and Myers~\cite{myers}.  From this perspective,
the expansion is due to the fact that the transverse coordinates
of the D$(p-2)$-branes are matrix-valued.  Myers' original analysis
was of a collection of $N$ D0-branes, in the presence of a background
R-R four-form field strength: they spontaneously expand into a D2-brane of
topology $\mathbb{R}
\times S^2_{NC}$, where $S^2_{NC}$ denotes the non-commutative two-sphere.
The D2-brane is uncharged with respect to the four-form, but has a
non-zero dipole moment.  Hence the name of a ``dielectric'' D2-brane.
Configurations for arbitrary $p$, and more general configurations
involving fuzzy cosets, were described in~\cite{trivedi:00}.

These two descriptions of the dielectric effect --- abelian and
non-abelian --- are of course dual to one another.  In the limit of
large $N$, the non-commutative nature of the fuzzy two-sphere is lost,
it becomes a smooth manifold, and the two descriptions do indeed agree.

In principle, all such dielectric branes should have a corresponding
supergravity solution describing the back-reaction of the brane on
spacetime.  Technical difficulties, however, have limited the analysis
of such solutions to that of $N$ D4-branes~\cite{costa:01,emparan:01}
or $N$ F-strings~\cite{emparan:01} expanding into a D6-brane with
topology $\mathbb{M}^5 \times S^2$ and $\mathbb{M}^2 \times S^5$
respectively.  In both cases, there is a stable and an unstable radius
of the dielectric sphere, the form of the effective potential matching
the abelian worldvolume analysis precisely~\cite{brecher:01,costa:01}.

There is thus much evidence, from both the worldvolume and
supergravity perspectives,  that the dielectric effect is not limited
to D-branes, but that dielectric F-strings also exist.  One can
then pose the following question: can the expansion of F-strings into
a D$p$-brane be described from the point of view of the strings
themselves?  It is the purpose of this paper to answer precisely this
question\footnote{A recent attempt to answer this question from the
perspective of Matrix string theory has already been
made~\cite{pedro}, although the results seem somewhat opaque to the authors and there
is little overlap with this work.}.

Since, from the strings' perspective, the dielectric effect should be
due to matrix-valued coordinates, we are led to a consideration of
Matrix string theory~\cite{motl:97,banks:97,DVV}.  Starting from the
Matrix theory action~\cite{CH,BFSS}, one compactifies on a circle, reinterprets
the
resulting theory as a $(1+1)$-dimensional super Yang-Mills theory on
the dual circle~\cite{taylor}, and then performs the so-called 9-11
flip, which implements a further S- and T-duality~\cite{DVV}.  It is
easy to see that the result is a $(1+1)$-dimensional super Yang-Mills
theory describing $N$ fundamental strings in the type IIA theory.  In
the weak string coupling limit, one recovers $N$ copies of the
Green-Schwarz action describing light-cone gauge string theory.

To discuss possible dielectric solutions of Matrix string theory, we
need to know how the R-R potentials couple to the worldvolume of the
(Matrix) string.  Since the Matrix theory action in a weakly curved
background is known~\cite{KT,TVR3}, one can in principle run through
the above chain of dualities to derive the Matrix string theory action
in a corresponding weakly curved background.  Indeed, Schiappa has
already considered this computation, and has written down an obvious
dielectric-like solution~\cite{schiappa}.  However, it is not at all
clear that this solution corresponds to a dielectric \emph{string},
for reasons explained  later in this paper.  Moreover, we find
couplings to various components of the R-R fields over and above those
found by Schiappa~\cite{schiappa} and, for this reason, we consider
the problem of deriving the Matrix string theory action in a weakly
curved background from scratch.  We should further note that one of
us has also already derived couplings of the type IIA F-string to various
background R-R potentials~\cite{yolanda}.

One might think that F-strings could expand into a D$p$-brane with
topology $\mathbb{M}^2 \times S^{p-1}$ for any
value of $p$, but this is not the case.  As mentioned above, one
generically finds a stable and an unstable spherical solution of the
D$p$-brane theory with dissolved electric flux.  Expansion into a
D2-brane, however, is atypical: only an \emph{unstable} solution
exists~\cite{emparan:97}.  Indeed, this observation is mirrored in the
corresponding (albeit smeared) supergravity
solution~\cite{brecher:01}.  One might suspect, therefore, that
stable cylindrical D2-brane solutions of the type IIA Matrix string theory do not
exist,
and we aim to address this issue here.

A further T-duality takes us to a type IIB Matrix string theory which,
as the S-dual of the D-string theory, describes $N$ F-strings in the
static gauge.  Just as D-strings can expand into D3-branes, so can
F-strings and the dielectric solution presented in~\cite{myers} is
equally applicable here.

In the following section, we consider Matrix theory in a weakly curved
background~\cite{KT,TVR3}.  It is easier to consider the chain of dualities
leading to Matrix string theory in ten- rather than eleven-dimensional
language, so we choose to work with the D0-brane theory in a weakly
curved background.  This has been derived from the Matrix theory
results in~\cite{TVR} --- we show explicitly that it reproduces the
lowest order expansion of the non-abelian D0-brane theory of Taylor
and van Raamsdonk~\cite{TVR2} and Myers~\cite{myers}.  T-duality,
taking us to the D-string theory, is considered in section 3, and
the 9-11 flip in section 4.  We show that the resulting linear
action reproduces the light-cone gauge string theory action in a
weakly curved background, thus lending some
weight to our results.  Indeed, we need the extra couplings, relative
to the results of~\cite{schiappa}, for this to be the case.  A further
T-duality taking us to the type IIB theory is considered in section 5
and, although we cannot perform the S-duality rigorously, we
argue that the result is equivalent to the S-dual of the D-string
theory.  In section 6, we apply once more a T-duality transformation,
this time in a direction transverse to the IIB string, giving a theory
of type IIA F-strings with winding number rather than light-cone momentum.  Some
of the
resulting couplings of both the type IIA and the type IIB strings have
already been considered by one of us~\cite{yolanda}.  Dielectric solutions
are considered in section 7, where we give some explicit solutions
describing the expansion of F-strings into D$p$-branes for different
$p$.  We conclude in section 8.

\sect{Matrix theory}

The bosonic sector of the Matrix theory action in a flat background
is~\cite{BFSS}
\be
S_{{\rm flat}} = \frac{1}{R} \int dt ~\Tr \left[ \frac{1}{2} D_t X^i D_t X^i +
\frac{R^2}{16\pi^2l_P^6} [X^i,
X^j][X^i,X^j] \right],
\label{eqn:matrix_action}
\ee

\noindent where $l_P$ denotes the eleven-dimensional Planck length,
$R$ is the radius of the eleventh dimension and $D_t X^i = \partial_t
X^i + i [A_t,X^i]$.  We will choose the gauge $A_t=0$ throughout this
paper at the expense of losing explicit gauge invariance.  With the
string coupling set through the relation $R=g_s \sqrt{\al'}$, this action is the
non-relativistic
limit of the non-abelian Born-Infeld action describing $N$ D0-branes, with mass
quantized in units of $1/R$: the dimensional reduction to one
dimension of ten-dimensional Yang-Mills theory with $g_{YM}^2 =
g_s/(4\pi^2\al'^{3/2})$.  The connections are more subtle,
however (see, e.g.,~\cite{polchinski:99}).  After all, when $N$ is large, Matrix
theory captures
eleven-dimensional physics in the infinite momentum frame~\cite{BFSS}.  For
finite $N$, it is equivalent to a DLCQ or null compactification of
M-theory~\cite{Suss,Sei,Sen}, whereas the type IIA D0-brane theory
comes about via a spacelike compactification.  The two actions are
then related by an infinite boost in the eleventh dimension.  At any
rate, if we take $T_0 = 1/R$ and $R = g_s \sqrt{\al'}$, so that $l_P
= g_s^{1/3} \sqrt{\al'}$, then we recover the Yang-Mills description of
D0-branes.

\subsection{Linear couplings in Matrix theory}

Matrix theory in an arbitrary background is understood only rather
poorly (although see~\cite{douglas,DKO,DO} for early work on this
subject) and, in the above
sense, this is related to the question as to what is the form of the
non-abelian Born-Infeld theory in curved space.  Kabat and Taylor have
derived the linear Matrix theory couplings to bosonic background
fields~\cite{KT} and Taylor and van Raamsdonk have extended these
calculations to include fermionic backgrounds
\cite{TVR3}.  They have further derived the linear couplings of the D0-brane~\cite{TVR} and
D$p$-brane~\cite{TVR2} theories from Matrix theory.  The results
certainly seem to agree with the linear order expansion of the
combined non-abelian Born-Infeld-Chern-Simons action proposed by Taylor and
van Raamsdonk~\cite{TVR2} and Myers~\cite{myers}, including the form
of the overall symmetrized trace due to Tseytlin~\cite{tseyt}.  More precise expressions for the Matrix
theory couplings, to all orders in both derivatives of the background
fields and the fermionic coordinates, can be had by dimensional reduction of
the eleven-dimensional supermembrane vertex operators constructed in~\cite{DNP,Plef}.

In other words, one can write the Matrix theory action as
\be
S = S_{{\rm flat}} + S_{{\rm linear}},
\ee

\noindent with $S_{{\rm flat}}$ given by (\ref{eqn:matrix_action}), and where
the linear action has the form~\cite{KT}
\be
S_{{\rm linear}} = \frac{1}{R} \int dt ~\STr \left\{ \frac{1}{2} h_{AB} \T^{AB}
+ A_{ABC} \J^{ABC} + A_{ABCDEF} \M^{ABCDEF} \right\},
\ee

\noindent where $A,B = 0,\ldots,10$ and $\STr$ denotes the symmetrized
trace\footnote{We include both the overall factor
of $1/R$ and the overall gauge trace in the action, rather than in the
currents, for convenience.}.  The eleven-dimensional metric,
$h_{AB}$, 3-form potential, $A_{ABC}$, and its 6-form dual,
$A_{ABCDEF}$, couple to the energy-momentum tensor, membrane current
and 5-brane current respectively.
The form of these currents has been worked out explicitly, and they match
eleven-dimensional
supergravity predictions~\cite{KT}.  There are also higher-order
multipole couplings to derivatives of the background fields, but we
will not be concerned with them here.  One
can relate this linear action to that relevant to the description of
D0-branes, via the infinite boost mentioned above.  The D0-brane
currents, denoted $I$, are related in this manner to the eleven-dimensional
currents $\T, \J$ and $\M$, as explained in~\cite{TVR}.

The linear D0-brane action is then
\bea
S_{{\rm linear}} &=& \frac{1}{R} \int dt ~\STr \left\{ \frac{1}{2} h_{\mn}
I_h^{\mn} +
\phi I_\phi + b_{\mn} I_s^{\mn} + \td{b}_{\mu_1 \ldots \mu_6} I_5^{\mu_1 \ldots
\mu_6} \right. \nn
&& \qquad \qquad \qquad \left. + C^{(1)}_{\mu} I_0^{\mu} + C^{(3)}_{\mnr}
I_2^{\mnr} + \frac{1}{60} C^{(5)}_{\mu_1 \ldots \mu_5} I_4^{\mu_1 \ldots \mu_5}
+ \frac{1}{336} C^{(7)}_{\mu_1 \ldots \mu_7} I_6^{\mu_1 \ldots \mu_7} \right\},
\label{eqn:D0}
\eea

\noindent where $\mu, \nu = 0,\ldots,9$.  The currents $I_h, I_s, I_5$ and
$I_p$ couple respectively to the metric, $h_{\mu \nu}$, the
Neveu-Schwarz-Neveu-Schwarz (NS-NS)
2-form potential $b^{(2)}$, its 6-form Hodge dual $\td{b}^{(6)}$, and
the R-R $(p+1)$-form potentials $C^{(p+1)}$.   The
potentials $C^{(5)}$ and $C^{(7)}$, the Hodge duals of $C^{(3)}$
and $C^{(1)}$, have been rescaled relative to~\cite{TVR,TVR2}.  Note
that there should also be couplings to $C^{(9)}$, but these are not
determined by the analysis of~\cite{TVR,TVR2}.  The currents
appearing in (\ref{eqn:D0}) are given in terms of the dimensional
reduction of the Born-Infeld field strength,
\be
F_{0i} = -F^{0i} = \partial_t X^i \equiv \dot{X}^i, \qquad F_{ij} = F^{ij} =
\frac{R}{2\pi l_P^3} i[X^i, X^j],
\label{eqn:BI}
\ee

\noindent where $i,j=1,\ldots,9$.  Usually $R$ is taken to be
$R=g_s\sqrt{\al'}$, and so one has $\la\equiv 2\pi \al' =2\pi l_P^3/R$.
After the 9-11 flip of section 4, however, the role of the ninth and
the eleventh coordinate are interchanged, and this relation is no
longer true.  For this reason we define a general quantity $\beta
\equiv 2\pi l_P^3/R$, the D0-brane theory and its duals being
recovered upon taking $\beta=\la$.

Substituting for (\ref{eqn:BI}), the NS-NS currents are~\cite{TVR}:
\bea
I_{\p} &=& \unity - \frac{1}{2} F^{0i}F^{0i} + \frac{1}{4} F^{ij}F^{ij} -
\frac{1}{8} \left( F^{\mu \nu}F_{\nu \rho}F^{\rho \sigma}F_{\sigma \mu} -
\frac{1}{4}
(F^{\mu \nu}F_{\mu \nu})^2 \right) \nn
&=& \unity - \frac{1}{4\beta^2} [X,X]^2 - \frac{1}{8\beta^4} \left(
[X,X]^4 - \quarter \left( [X,X]^2 \right)^2
\right) \nn
&& \qquad \qquad \qquad \qquad \qquad \qquad \qquad \qquad - \frac{1}{2}
\dot{X}^2 \left( \unity + \frac{1}{4}
\dot{X}^2 - \frac{1}{4\beta^2} [X,X]^2 \right),
\label{eqn:dilaton_current} \\
I_h^{00} &=& \unity + \frac{1}{2} F^{0i}F^{0i} +
\frac{1}{4}F^{ij}F^{ij} = \unity + \frac{1}{2} \dot{X}^2 -
\frac{1}{4\beta^2} [X,X]^2, \label{eqn:h_current} \\
I_h^{0i} &=& -F^{0i} \left( \unity +\frac{1}{2} F^{0j}F^{0j} +
\frac{1}{4} F^{jk}F^{jk} \right) - F^{ij}F^{jk}F^{0k} \nn
&=& \dot{X}^i \left( \unity + \frac{1}{2} \dot{X}^2 -
\frac{1}{4\beta^2} [X,X]^2 \right) -\frac{1}{\beta^2}
[X^i,X^j][X^j,X^k]\dot{X}^k, \\
I_h^{ij} &=& F^{0i}F^{0j} + F^{ik}F^{kj} = \dot{X}^i \dot{X}^j -
\frac{1}{\beta^2}  [X^i,X^k][X^k,X^j], \label{eqn:h_current2} \\
I_s^{0i} &=& \frac{1}{2} F^{ij}F^{0j} = - \frac{i}{2\beta} [X^i,X^j] \dot{X}^j,
\\
I_s^{ij} &=& \frac{1}{2} F^{ij} \left( \unity + \frac{1}{4}
F^{kl}F^{kl} - \frac{1}{2} F^{0j}F^{0j} \right) + \frac{1}{2} F^{0i}
F^{0k} F^{kj} - \frac{1}{2} F^{0j} F^{0k} F^{ki} + \frac{1}{2}
F^{ik}F^{kl}F^{lj} \nn
&=& \frac{i}{2\beta} [X^i,X^j] \left( \unity - \frac{1}{2}
\dot{X}^2 - \frac{1}{4\beta^2} [X,X]^2 \right) - \frac{i}{2\beta} \left(
\dot{X}^i
[X^j,X^k] \dot{X}^k - \dot{X}^j [X^i,X^k] \dot{X}^k \right) \nn
&& \qquad \qquad \qquad \qquad \qquad \qquad \qquad \qquad -
\frac{i}{2\beta^3} [X^i,X^k][X^k,X^l][X^l,X^j],
\label{eqn:b_current}
\eea
where we have defined
\bea
\dot{X}^2 &\equiv& \dot{X}^i \dot{X}^i, \nn
{[}X,X{]}^2 &\equiv&  [X^i,X^j][X^i,X^j],\nn
{[}X,X{]}^4 & \equiv & [X^i,X^j][X^j,X^k][X^k,X^l][X^l,X^i]. \nonumber
\eea
The R-R currents are~\cite{TVR}
\bea
I_0^0 &=& \unity, \label{eqn:rr1} \\
I_0^i &=& - F^{0i} = \dot{X}^i, \\
I_2^{0ij} &=& -\frac{1}{6} F^{ij} = - \frac{i}{6\beta} [X^i,X^j], \\
I_2^{ijk} &=& \frac{1}{6} \left( F^{0i}F^{jk} + F^{0j}F^{ki} +
F^{0k}F^{ij} \right) \nn
&=& - \frac{i}{6\beta} \left( \dot{X}^i [X^j,X^k] +
\dot{X}^j [X^k,X^i] + \dot{X}^k [X^i,X^j] \right), \\
I_4^{0ijkl} &=& \frac{1}{2} \left( F^{ij}F^{kl} + F^{ik}F^{lj} +
F^{il}F^{jk} \right) \nn
&=& - \frac{1}{2\beta^2} \left( [X^i,X^j][X^k,X^l] +
[X^i,X^k][X^l,X^j] + [X^i,X^l][X^j,X^k] \right), \\
I_4^{ijklm} &=& -\frac{15}{2} F^{0[i} F^{jk} F^{lm]} = -
\frac{15}{2\beta^2} \dot{X}^{[i} [X^j,X^k] [X^l,X^{m]}], \\
I_6^{0ijklmn} &=& - F^{[ij}F^{kl}F^{mn]} = \frac{i}{\beta^3}
[X^{[i},X^j] [X^k,X^l] [X^m,X^{n]}], \\
I_6^{ijklmnp} &=& 7 F^{0[i}F^{jk}F^{lm}F^{np]} = \frac{7i}{\beta^3}
\dot{X}^{[i} [X^j,X^k] [X^l,X^m] [X^n,X^{p]}].
\label{eqn:rr2}
\eea

Up to a couple of caveats, we show below that the above currents agree with the
expansion, to
linear order in the background fields, of the multiple D0-brane theory
of Taylor and van Raamsdonk~\cite{TVR2} and Myers~\cite{myers}.  The first such
caveat is
that we must have $I_5^{\mu_1 \ldots \mu_6} = 0$ if the two actions are to
match~\cite{myers}.  In other words, D0-branes cannot
expand into NS5-branes.  We will therefore drop this term from now
on.  The second is that the currents $I_h^{00}$ and $I_h^{ij}$ derived
from Matrix theory actually contain
further terms~\cite{TVR}, the form of which seems less certain.
Indeed, although the expressions (\ref{eqn:h_current}) and
(\ref{eqn:h_current2}) match precisely the first-order
expansion of the non-abelian Born-Infeld action, as given
in~\cite{TVR2, myers}, we have not been able to match these extra terms in
$I_h^{00}$ and $I_h^{ij}$.  We will, for this reason, ignore them.

\subsection{Matrix theory vs. multiple D0-branes}

We will now show that the above currents are recovered in the expansion to linear
order in the background fields of the multiple D0-brane theory\footnote{See also \cite{hyak}.}.
If we set the NS-NS fields to zero, the multiple D0-brane action is~\cite{tseyt,TVR2,
myers}
\bea
S &=& S_{{\rm flat}} + S_{{\rm CS}}, \label{eqn:d0-action} \\
S_{{\rm flat}} &=& -T_0 \int dt ~\STr \left\{ \sqrt{ (1 - \dot{X}^i
(Q\inv )_{ij} \dot{X}^j ) \det (Q_{ij}) } \right\},
\label{eqn:nbi_flat} \\
S_{{\rm CS}} &=& T_0 \int \STr \left\{ P \left[ e^{ i (\incl_X
\incl_X)/\la} \left( \sum C^{(n)} \right) \right] \right\},
\label{eqn:chern-simons}
\eea

\noindent where $T_0 = 1/R = 1/(g_s\sqrt{\al'})$ and
\be
Q^{ij} = \delta^{ij} + \frac{i}{\la} [X^i, X^j].
\ee

\noindent The pull-back is defined in terms of gauge covariant
derivatives, such that
\be
P[ \omega ] = (\omega_0 + D_tX^i \omega_i)~dt = (\omega_0 + \dot{X}^i
\omega_i)~dt,
\ee

\noindent with obvious generalizations to forms of higher degree.  The interior
multiplication
is
\be
( \incl_X \Sigma )_{\mu_1 \ldots \mu_{p}} = X^i \Sigma_{i \mu_1 \ldots
\mu_{p}},
\ee
\noindent giving rise to the relevant commutators in the action.
The Chern-Simons action (\ref{eqn:chern-simons}) becomes
\bea
S_{{\rm CS}} &=& T_0 \int \STr \left\{ P \left[ C^{(1)} + \frac{i}{\la}
\left(\incl_X \incl_X \right) C^{(3)} - \frac{1}{2\la^2} \left(\incl_X \incl_X
\right)^2 C^{(5)} - \frac{i}{6\la^3} \left(\incl_X \incl_X
\right)^3 C^{(7)} \right.\right. \nonumber\\
&&\left.\left. \qquad \qquad \qquad \qquad \qquad + \frac{1}{24\la^4}
\left(\incl_X \incl_X \right)^4
C^{(9)}\right] \right\}.
\label{eqn:cs}
\eea

\noindent It is easy to see that the D0-brane R-R currents,
(\ref{eqn:rr1})-(\ref{eqn:rr2}), of the previous subsection, can be written such
that
\bea
C^{(1)}_\mu I_0^\mu ~ dt &=& P[ C^{(1)} ], \\
C^{(3)}_{\mnr} I_2^{\mnr} ~ dt &=& \frac{i}{\la} P[ \incl_X \incl_X
C^{(3)} ], \\
\frac{1}{60} C^{(5)}_{\mu_1 \ldots \mu_5} I_4^{\mu_1 \ldots \mu_5} ~ dt &=&
-\frac{1}{2\la^2} P[ \left( \incl_X \incl_X \right)^2 C^{(5)} ], \\
\frac{1}{336} C^{(7)}_{\mu_1 \ldots \mu_7} I_6^{\mu_1 \ldots \mu_7} ~ dt &=&
-\frac{i}{6\la^3} P[ \left( \incl_X \incl_X \right)^3 C^{(7)} ],
\eea

\noindent which reproduce the Chern-Simons action (\ref{eqn:cs}).
Since we have set the NS-NS 2-form to zero, these linear couplings are actually
exact.  We should also note that one could derive the correct Matrix theory
coupling to $C^{(9)}$ in (\ref{eqn:D0}) by comparing with the
Chern-Simons action (\ref{eqn:cs}).

The NS-NS fields are somewhat harder to deal with.  To linear order
in the background fields, the non-abelian Born-Infeld action is
\be
S_{{\rm NBI}} = S_{{\rm NBI}}(\p=0) - \p S_{{\rm flat}},
\ee

\noindent with $S_{{\rm flat}}$ as in (\ref{eqn:nbi_flat}).  This gives the
dilaton current:
\be
I_{\p} = \sqrt{ (1 - \dot{X} Q\inv \dot{X} ) \det Q },
\label{eqn:nbi_dilaton}
\ee

\noindent where $\dot{X} Q\inv \dot{X} = \dot{X}^i
(Q\inv )_{ij} \dot{X}^j$.  As shown already in \cite{tseyt, myers},
expanding the current (\ref{eqn:nbi_dilaton}) to the
relevant order, gives the Matrix theory result
(\ref{eqn:dilaton_current}).

The action containing the couplings to the metric and NS-NS 2-form is
more complicated:
\be
S_{{\rm NBI}}(\p=0) = -T_0 \int dt ~\STr \left\{ \sqrt{ -P \left[ E_{00} +
E_{0i} ( \Q \inv - \delta )^i_{~k} E^{kj} E_{j0} \right] \det ( \Q^i_{~j} ) }
\right\},
\ee

\noindent where
\be
\Q^i_{~j} = \delta^i_{~j} + \frac{i}{\la} [X^i,X^k] E_{kj}.
\ee

\noindent To linear order in the background fields, $E_{\mn} =
\eta_{\mn} + h_{\mn} + b_{\mn}$, so that
\be
\Q^i_{~j} = Q^i_{~k} \left( \delta^k_{~j} + (Q\inv)^k_{~l} \frac{i}{\la}
[X^l,X^m] ( h_{mj} + b_{mj} ) \right),
\ee

\noindent which gives, again to linear order,
\be
(\Q\inv)^i_{~j} = \left( \delta^i_{~k} - (Q\inv)^i_{~l} \frac{i}{\la}
[X^l,X^m] ( h_{mk} + b_{mk} ) \right) (Q\inv)^k_{~j},
\ee

\noindent so we have all the ingredients we need to derive the
linear Born-Infeld couplings to the metric and NS-NS 2-form potential.  We find
\bea
I_h^{00} &=& \sqrt{\det Q} (1 - \dot{X} Q\inv \dot{X} )^{-1/2}, \\
I_h^{0i} &=& \sqrt{\det Q} (1 - \dot{X} Q\inv \dot{X} )^{-1/2} ~(Q\inv)^{(ij)}
\dot{X}^j, \\
I_h^{ij} &=& \frac{1}{2} \sqrt{\det Q} (1 - \dot{X} Q\inv \dot{X}
)^{-1/2} \left[ \dot{X}^i (Q\inv)^j_{~k} \dot{X}^k -
\frac{i}{\la} \dot{X}^l (Q\inv)_{lm} [X^m,X^i] (Q\inv)^j_{~k}
\dot{X}^k  \right. \nn
&& \qquad \qquad \left. - \frac{i}{\la} ( 1 - \dot{X} Q\inv \dot{X} )
(Q\inv)^j_{~k} [X^k,X^i] + (i \longleftrightarrow j) \right], \\
I_s^{0i} &=& \frac{1}{2} \sqrt{\det Q} (1 - \dot{X} Q\inv \dot{X}
)^{-1/2} ~(Q\inv)^{[ij]} \dot{X}^j, \\
I_s^{ij} &=& \quarter \sqrt{\det Q} (1 - \dot{X} Q\inv \dot{X}
)^{-1/2} \left[ \dot{X}^i (Q\inv)^j_{~k} \dot{X}^k -
\frac{i}{\la} \dot{X}^l (Q\inv)_{lm} [X^m,X^i] (Q\inv)^j_{~k}
\dot{X}^k \right. \nn
&& \qquad \qquad \left. - \frac{i}{\la} ( 1 - \dot{X} Q\inv \dot{X} )
(Q\inv)^j_{~k} [X^k,X^i] - (i \longleftrightarrow j) \right].
\eea

\noindent One can verify that the lowest order expansion of these
currents reproduces the results
(\ref{eqn:h_current})-(\ref{eqn:b_current}).

\sect{T-duality: multiple D-strings}

To construct the Matrix string theory of Dijkgraaf, Verlinde and
Verlinde~\cite{DVV}, one compactifies Matrix theory on a
circle in, say, the $x^9$ direction.  Taylor has shown that this is
equivalent to a $(1+1)$-dimensional super Yang-Mills theory on
the dual circle~\cite{taylor}.  Taking now $i,j = 1, \ldots,8$, and
denoting the dual coordinate by $\hat{x}$, the worldvolume fields
transform as~\cite{taylor}
\bea
F_{ij} = F^{ij} = \frac{i}{\beta} [X^i,X^j] &\longrightarrow& \frac{1}{2\pi
\hat{R}_9} \int d\hat{x} \frac{i}{\beta}[X^i,X^j], \label{eqn:t-duality1} \\
F_{9i} = F^{9i} = \frac{i}{\beta} [X^9,X^i] &\longrightarrow& \frac{1}{2\pi
\hat{R}_9} \int d\hat{x} \frac{\la}{\beta} D_{\hat{x}} X^i, \\
F_{0i} = -F^{0i} = \dot{X}^i &\longrightarrow& \frac{1}{2\pi \hat{R}_9} \int
d\hat{x}
\dot{X}^i, \\
F_{09} = -F^{09} = \dot{X}^9 &\longrightarrow& \frac{1}{2\pi \hat{R}_9} \int
d\hat{x} \la
\dot{A}_{\hat{x}},
\label{eqn:t-duality2}
\eea

\noindent where $\hat{R}_9 = \al'/R_9$ is the radius of the dual circle.  Of
course, this is just T-duality applied to
Matrix theory.  By construction, the multiple D0-brane action
(\ref{eqn:d0-action})
considered in the previous section is covariant
under T-duality, so an application of T-duality to the D0-brane action
(\ref{eqn:D0}) derived from Matrix theory should reproduce
the non-relativistic limit of the multiple D-string action.

Certainly, the action (\ref{eqn:matrix_action}) in a
flat background becomes, with $DX^i \equiv D_{\hat{x}} X^i$ and $A
\equiv A_{\hat{x}}$,
\be
S_{\rm flat} = \frac{1}{2\pi R \hat{R}_9} \int dt d\hat{x} ~\Tr \left[
\frac{1}{2} \dot{X}^2 - \frac{\la^2}{2\beta^2} DX^2 + \frac{\lambda^2}{2}
\dot{A}^2 +
\frac{1}{4\beta^2} [X^i, X^j]^2 \right],
\label{eqn:d1_flat}
\ee

\noindent which, with $\beta=\la$, is just the non-relativistic limit of the
flat space D-string action; with T-duality acting on the string coupling as
\be
g_s \longrightarrow \hat{g}_s = g_s \frac{\sqrt{\al'}}{R_9},
\ee

\noindent the overall factor is $T_1 = 1/(\hat{g}_s \la)$ as
required.  Turning to the linear action (\ref{eqn:D0}), we must consider
T-duality applied to both the worldvolume and background fields.  As far as the
currents are
concerned, we simply take the $I$s written in terms of
the Born-Infeld field strength (\ref{eqn:BI}), and reinterpret the relevant
components through the action of T-duality as given by
(\ref{eqn:t-duality1})-(\ref{eqn:t-duality2}) above~\cite{TVR2}.  This
is a simple re-writing, the results being collected in the
appendix.

To linear order, the action of T-duality on the background fields is:
\be
h_{a9} \longleftrightarrow -b_{a9}, \qquad h_{9a} \longleftrightarrow b_{9a},
\qquad h_{99} \longleftrightarrow -h_{99}, \qquad \p \longrightarrow
\p - \half h_{99},
\label{eqn:t-duality_bgd1}
\ee
\be
C^{(p)}_{a_1 \ldots a_{p-1} 9} \longleftrightarrow C^{(p-1)}_{a_1 \ldots
a_{p-1}},
\label{eqn:t-duality_bgd2}
\ee

\noindent where $a,b = 0, \ldots, 8$, and all other fields are
invariant.  A simple application of these rules gives
\bea
S_{{\rm linear}} &=& \frac{1}{2\pi R \hat{R}_9} \int dt d\hat{x} ~\STr \left\{
\frac{1}{2}
h_{ab} I_h^{ab} - 2 h_{a9} I_s^{a9} - \frac{1}{2} h_{99}I_h^{99} + \left
( \phi - \frac{1}{2} h_{99} \right) I_\phi - b_{a9} I_h^{a9}  + b_{ab}
I_s^{ab} \right.\nn
&&  \qquad
+ C^{(0)} I_0^9 + C^{(2)}_{a9} I_0^a + 3 C^{(2)}_{ab} I_2^{ab9} + C^{(4)}_{abc9}
I_2^{abc} + \frac{1}{12} C^{(4)}_{a_1 \ldots a_4} I_4^{a_1 \ldots a_4 9} \nn
&&\left. \qquad \qquad \qquad
+ \frac{1}{60} C^{(6)}_{a_1 \ldots a_5 9} I_4^{a_1
\ldots a_5}
+ \frac{1}{48} C^{(6)}_{a_1 \ldots a_6} I_6^{a_1 \ldots a_6 9}
+ \frac{1}{336} C^{(8)}_{a_1 \ldots a_7 9} I_6^{a_1 \ldots a_7}
\right\}.
\label{eqn:d1_linear}
\eea
There should also be an extra coupling to $C^{(8)}$, coming
from the unknown Matrix theory coupling to $C^{(9)}$.

The resulting action should be equivalent to the linearized
version of the D-string action.  With the results in the appendix, it is
easy to see that the R-R terms can indeed be written as
\[
S_{{\rm R-R}} = T_1 \int \STr \left\{ P \left[ \la C^{(0)}
\wedge F + C^{(2)} + i (\incl_X \incl_X) C^{(2)} \wedge F +
\frac{i}{\la} (\incl_X \incl_X) C^{(4)} - \frac{1}{2\la} (\incl_X
\incl_X)^2 C^{(4)} \wedge F \right. \right.
\]
\be
\left. \left. - \frac{1}{2\la^2} (\incl_X \incl_X)^2
C^{(6)} - \frac{i}{6\la^2} (\incl_X \incl_X)^3 C^{(6)} \wedge F -
\frac{i}{6\la^3} (\incl_X \incl_X)^3 C^{(8)} + \frac{1}{24\la^3}
(\incl_X \incl_X)^4 C^{(8)} \wedge F \right] \right\},
\label{eqn:d1_pullbacks}
\ee
where $F_{09} = \dot{A}$.  We should note that only half of
the terms necessary to form the pullback of $C^{(8)}$ are present in
the linear action (\ref{eqn:d1_linear}).  The missing
terms come from the $C^{(9)}$ coupling in the D0-brane action (\ref{eqn:D0}), as
mentioned above.  At any rate, this is just the expansion of the
Chern-Simons action
\be
S_{{\rm CS}} = T_1 \int \STr \left\{ P \left[ e^{ i (\incl_X
\incl_X)/\la} \left( \sum C^{(n)} \right) \right] \wedge e^{\la F} \right\},
\ee
as expected.  It is computationally more involved to check the
NS-NS couplings in (\ref{eqn:d1_linear}) against the non-abelian
Born-Infeld theory of D-strings, and we will not do this here.

\sect{Matrix string theory: multiple IIA F-strings}

\subsection{The 9-11 flip}
\label{9-11}

Having constructed the $(1+1)$-dimensional theory of the D-string, we
are now in a position to perform the so-called 9-11 flip, a rotation
\be
x^9 \longrightarrow x^{11}, \qquad x^{11} \longrightarrow -x^9,
\ee
which will give us the type IIA Matrix string
theory action in a linear background.  Whereas
Schiappa~\cite{schiappa} considers the action of the 9-11 flip
on the currents, and leaves the background fields invariant, we take
the view here that it is the currents which are invariant under the
9-11 flip.  After all, in the flat space case, the 9-11 flip does not
change the worldvolume fields~\cite{DVV}.  Moreover, Schiappa has
argued that the currents are in fact
invariant under the S- and T-dualities~\cite{schiappa}. We simply take $R_9 = g_s
\sqrt{\al'}$, so that $\hat{R}_9 = \sqrt{\al'}/g_s$ and $l_P =
g_s^{4/3}\hat{R}_9$.  We then define the dimensionless worldsheet coordinates
\be
\sigma = \frac{\hat{x}}{\hat{R}_9}, \qquad \tau = \frac{R}{\al'} t,
\label{eqn:rescalings}
\ee

\noindent and perform the rescalings $X^i
\longrightarrow \sqrt{\al'} X^i$ and $g_s \longrightarrow
g_s/(2\pi)$.  The flat space action (\ref{eqn:d1_flat}) becomes
the Matrix string action of \cite{DVV}:
\be
S_{\rm flat} = \frac{1}{2\pi} \int d\tau d\sigma ~\Tr \left[
\frac{1}{2} \dot{X}^2 - \frac{1}{2} DX^2 + \frac{g_s^2}{2} \dot{A}^2 +
\frac{1}{4g_s^2} [X, X]^2 \right],
\label{eqn:flat_mst}
\ee

\noindent where $-\infty < \tau < \infty$ and $0 \le \sigma < 2\pi$.
Weakly coupled string theory at $g_s=0$ is recovered in the IR limit,
so is described by strongly coupled $(1+1)$-dimensional Yang-Mills.  The
conformal field theory which describes this IR limit is a
sigma model on an orbifold target space~\cite{DVV}.  The matrix-valued
coordinates must commute in this limit, so can be simultaneously
diagonalised, the eigenvalues $x^i_1, \ldots,
x^i_N$ corresponding to the positions of the $N$ strings.  Then the action
(\ref{eqn:flat_mst}) reduces to a sum of Green-Schwarz actions for
light-cone gauge string theory:
\be
S_{\rm flat} = \frac{1}{2\pi} \int d\tau d\sigma \sum_{n=1}^N \left[
\frac{1}{2} \dot{x}_n^2 - \frac{1}{2} \del x_n^2 \right].
\ee

\noindent The couplings to linear background
fields considered herein should thus tell us something about light-cone gauge
string
theory in weakly curved backgrounds.

The action of the 9-11 flip on the background
fields is easy to derive.  Consider, for example, the type IIB metric
fluctuation  $h_{a9}$. Its T-dual on the type IIA side
is $-b_{a9}$, or in eleven-dimensional language $-A_{a9\, 11}$. Under the 9-11
flip $-A_{a9\, 11} \rightarrow A_{a11\,9}=-b_{a9}$.  In
other words, $h_{a9} \rightarrow -b_{a9}$ under the 9-11 flip, which is just
(linearized) S-duality followed by (linearized) T-duality in the $x^9$
direction.  Arguing in a similar manner, one finds that the linear background
fields transform in the following way under the 9-11 flip:
\be
\begin{array}{lll}
h_{ab} \longrightarrow h_{ab} - \eta_{ab} \left( \p -\frac{1}{2}h_{99} \right),
& h_{a9} \longrightarrow -b_{a9},
&h_{99} \longrightarrow -\left(\p + \frac{1}{2}h_{99} \right),\label{eqn:911_h}
\\ \\
\p \longrightarrow -\p + \frac{1}{2} h_{99},
& b_{ab} \longrightarrow -C^{(3)}_{ab9},
&b_{a9} \longrightarrow-C^{(1)}_a,
\\ \\
C^{(0)} \longrightarrow -C^{(1)}_9,
& C^{(2)}_{ab} \longrightarrow b_{ab},
& C^{(2)}_{a9} \longrightarrow -h_{a9},
\\ \\
C^{(4)}_{a_1 \ldots a_4} \longrightarrow C^{(5)}_{a_1 \ldots a_49},
&C^{(4)}_{abc9} \longrightarrow C^{(3)}_{abc},
& C^{(6)}_{a_1\ldots a_6} \longrightarrow N^{(7)}_{a_1 \ldots a_6 9},
\\ \\
C^{(6)}_{a_1 \ldots a_5 9} \longrightarrow {\tilde b}_{a_1 \ldots a_5 9},
&C^{(8)}_{a_1 \ldots a_8} \longrightarrow -C^{(9)}_{a_1 \ldots a_8 9},
& C^{(8)}_{a_1 \ldots a_7 9} \longrightarrow -C^{(7)}_{a_1 \ldots
a_7}.
\end{array}
\ee
Here, $N^{(7)}_{a_1 \ldots a_6 9}$ is the field that couples
minimally to a type IIA Kaluza-Klein monopole whose Taub-NUT direction
is along $x^9$.  It is easy to verify that the above
transformations are precisely what one would find by performing a linearized
S-duality (in the string frame)
\be
h \longrightarrow h - \eta \p, \qquad \p
\longrightarrow -\p, \qquad b_2 \longrightarrow -C^{(2)},
\label{S-rules1}
\ee
\be
C^{(0)} \longrightarrow -C^{(0)}, \qquad C^{(2)} \longrightarrow b_2,
\qquad C^{(6)} \longrightarrow \td{b}_6, \qquad C^{(8)}
\longrightarrow -C^{(8)},
\label{S-rules2}
\ee

\noindent followed by a linearized T-duality in the $x^9$ direction,
as in (\ref{eqn:t-duality_bgd1}) and (\ref{eqn:t-duality_bgd2}),
plus the linearized T-duality rules for the field ${\tilde b}_6$ \cite{EJL}:

\be
\label{T-rules3}
{\tilde b}_{a_1 \ldots a_6} \longrightarrow N^{(7)}_{a_1 \ldots a_6 9},
\qquad \qquad {\tilde b}_{a_1 \ldots a_5 9} \longrightarrow
{\tilde b}_{a_1 \ldots a_5 9}.
\ee

Performing the transformations (\ref{eqn:911_h}) on
the linear action (\ref{eqn:d1_linear}), and substituting for
(\ref{eqn:rescalings}), one finds
\[
S_{{\rm linear}}^{{\rm IIA}} = \frac{1}{2\pi} \int d\tau d\sigma \frac{\al'}{R^2}
~\STr \left\{
\frac{1}{2}
\left( h_{ab} - \eta_{ab} \left(\p - \frac{1}{2} h_{99} \right)
\right) I_h^{ab} + 2 b_{a9} I_s^{a9} + \frac{1}{2} \left(\p + \frac{1}{2} h_{99}
\right) I_h^{99} \right.
\]
\[
- \frac{1}{2} \left( \p - \frac{3}{2} h_{99} \right) I_\phi + C^{(1)}_a I_h^{a9}
- C^{(3)}_{ab9}
I_s^{ab} - C^{(1)}_9 I_0^9 - h_{a9} I_0^a + 3 b_{ab} I_2^{ab9} + C^{(3)}_{abc}
I_2^{abc}
\]
\be
\left.
 + \frac{1}{12} C^{(5)}_{a_1 \ldots a_4 9} I_4^{a_1 \ldots a_4 9}
+ \frac{1}{60} \td{b}_{a_1 \ldots a_5 9} I_4^{a_1 \ldots a_5}
+ \frac{1}{48} N^{(7)}_{a_1 \ldots a_69} I_6^{a_1 \ldots a_6 9}
- \frac{1}{336} C^{(7)}_{a_1 \ldots a_7} I_6^{a_1 \ldots a_7}
\right\}.
\label{eqn:linear_mst}
\ee

\noindent Writing the currents in the appendix in terms of the dimensionless
quantities $\tau$ and
$\sigma$, and after rescaling $X^i \longrightarrow \sqrt{\al'} X^i$ and $g_s
\longrightarrow g_s/(2\pi)$, this is the action describing Matrix
string theory in a weakly curved background.  There is no need to
write out the new couplings in full, suffice it to say that each term
of the form $\dot{X}, \la DX/\beta, [X,X]/\beta$ and
$\la \dot{A}$ appears multiplied by a factor of $R/\sqrt{\al'}$, each
$[X,X]$ term appears multiplied by a factor of $1/g_s$, and each $\dot{A}$
term appears with a factor of $g_s$.  We should note that the
couplings derived by Schiappa~\cite{schiappa} are as above, but
\emph{without} those couplings to fields with a component in the $x^9$
direction.  As we will see below, these latter are necessary to match
with light-cone gauge string theory.

\subsection{Light-cone gauge string theory in a general background}

To examine the action (\ref{eqn:linear_mst}), let us consider the weakly coupled
string theory, $g_s = 0$.  Just as in the flat case, the Born-Infeld field
strength drops out entirely, and one is forced onto the space of
commuting matrices as above.  Moreover, the couplings to all Ramond-Ramond
fields vanish, as one might expect, and we find
\[
S_{{\rm linear}}^{{\rm IIA}} = \frac{1}{2\pi} \int d\tau d\sigma ~ \sum_{n=1}^N
\left\{
\frac{\al'}{R^2} \left( \frac{1}{2} h_{00} - h_{09} + \frac{1}{2}h_{99} \right)
+ \frac{\sqrt{\al'}}{R} \left( \vphantom{\frac{1}{2}} (h_{0i} - h_{9i})
\dot{x}^i_n + (b_{0i} -
b_{9i}) \del x^i_n \right) \right.
\]
\[
+ \frac{1}{4} (h_{00} - h_{99}) (\dot{x}_n^2 + \del x_n^2) +
\frac{1}{2} h_{ij} (\dot{x}^i_n \dot{x}^j_n - \del x^i_n \del x^j_n)
+ b_{ij} \dot{x}_n^i \del x_n^j + b_{09} \dot{x}_n \cdot \del x_n
\]
\[
+ \frac{R}{\sqrt{\al'}} \left( \frac{1}{2} h_{0i} \dot{x}_n^i
(\dot{x}_n^2 + \del x_n^2)
+ b_{i9} \left( \frac{1}{2} \del x_n^i (\dot{x}_n^2 + \del x_n^2) +
\dot{x}_n^i \del x_n \cdot \dot{x}_n \right) \right)
\]
\be
+ \left. \frac{R^2}{\al'} \left( \frac{1}{4} \p -\frac{3}{8} h_{99} \right)
\left( (\dot{x}_n \cdot \del x_n)^2 + \frac{1}{4} (\dot{x}_n^2 + \del x_n^2)^2
\right) \vphantom{\sum_{n=1}^N} \right\}.
\ee

\noindent If we now define the (somewhat non-standard) light-cone coordinates
\be
X^{\pm} = \frac{1}{\sqrt{2}} \left( X^0 \mp X^9 \right),
\ee

\noindent then we have
\[
S_{{\rm linear}}^{{\rm IIA}} = \frac{1}{2\pi} \int d\tau d\sigma ~ \sum_{n=1}^N
\left\{
\frac{\al'}{R^2} h_{++} + \frac{\sqrt{\al'}}{R} \sqrt{2} \left(
h_{+i}\dot{x}_n^i + b_{+i} \del x_n^i \right) + \frac{1}{2} h_{+-}
(\dot{x}_n^2 + \del x_n^2) \right.
\]
\be
\left. + \frac{1}{2} h_{ij} ( \dot{x}_n^i \dot{x}_n^j - \del x_n^i
\del x_n^j) + b_{ij} \dot{x}_n^i \del x_n^j + b_{+-} \dot{x}_n \cdot
\del x_n + \cO \left( \frac{R}{\sqrt{\al'}} \right) \right\}.
\label{eqn:lc_mst}
\ee

We wish to compare this action to light-cone gauge string theory in a weakly
curved background.  To this end, consider
\be
S = -\frac{1}{4\pi \al'} \int d\tau d\sigma \left( \sqrt{-\gamma}
\gamma^{\alpha \beta} \del_{\alpha} x^\mu \del_{\beta} x^\nu G_{\mn} (x) +
\epsilon^{\alpha
\beta} \del_{\alpha} x^\mu \del_{\beta} x^\nu B_{\mn}(x) \right),
\label{eqn:lc_string}
\ee

\noindent where $\alpha, \beta$ denote the worldsheet coordinates $\tau$ and
$\sigma$, $\gamma_{\alpha\beta}$ is the worldsheet metric and $\gamma$ its
determinant.  We are always free to take the worldsheet to be flat, in which case
\be
S = -\frac{1}{4\pi \al'} \int d\tau d\sigma \left(\vphantom{\frac{1}{2}}
-\dot{x}^\mu
\dot{x}^\nu G_{\mn} (x) + \del x^\mu \del x^\nu G_{\mn} (x)
- 2 \dot{x}^\mu \del x^\nu B_{\mn} (x) \right).
\ee

\noindent Since the worldsheet energy-momentum tensor vanishes,
this must be supplemented with the constraint
\be
(\dot{x}^\mu \pm \del x^\mu)(\dot{x}^\nu \pm \del x^\nu) G_{\mn} = 0.
\label{eqn:constraint}
\ee

The light-cone gauge is defined by taking $x^+(\tau,\sigma) = \tau$.  Since we
interested only in linear backgrounds, we set $G_{\mn} = \eta_{\mn} +
h_{\mn}$, where $\eta_{+-} = -1$ and $\eta_{ij} = \delta_{ij}$, and
$B_{\mn} = b_{\mn}$.  As usual in a light-cone treatment of
gravity~\cite{GS}, we can further use spacetime
diffeomorphisms to set $G_{--} = 0 = G_{-i}$.  The string action is then
\[
S = \frac{1}{2\pi \al'} \int d\tau d\sigma \left\{ -2\dot{x}^- +
\dot{x}^2 - \del x^2 + \frac{1}{2} h_{++} + h_{+i} \dot{x}^i + h_{+-}
\dot{x}^- + \frac{1}{2} h_{ij} (\dot{x}^i \dot{x}^j - \del x^i \del
x^j) \right.
\]
\be
\left. + b_{+-} \del x^-  + b_{+i} \del x^i + b_{ij} \dot{x}^i
\del x^j + b_{-i} \left( \frac{1}{2} (\dot{x}^2 + \del x^2 )
\del x^i - \dot{x} \cdot \del x \dot{x}^i \right) \right\}.
\ee

\noindent At first sight, the linear couplings to the background
fields do not seem to match those in the action (\ref{eqn:lc_mst}).
However, the constraint (\ref{eqn:constraint}) can be
solved for $x^-$ giving, again to linear order in $h_{\mn}$,
\bea
\dot{x}^- &=& \frac{1}{2} (\dot{x}^2 + \del x^2 ) + \frac{1}{2} h_{++}
+ \dot{x}^i h_{+i} + \frac{1}{2} (\dot{x}^i \dot{x}^j + \del x^i \del
x^j ) h_{ij} + \frac{1}{2} (\dot{x}^2 + \del x^2 ) h_{+-},
\label{eqn:constraint1}\\
\del x^- &=& \dot{x} \cdot \del x + \del x^i h_{+i} + \dot{x}^i \del
x^j h_{ij} + \dot{x} \cdot \del x h_{+-}.
\label{eqn:constraint2}
\eea

\noindent To linear order, then, we can replace $\dot{x}^-$ and $\del
x^-$ in the string action with $(\dot{x}^2 + \del x^2 )/2$ and
$\dot{x} \cdot \del x$ respectively.  In that case, we find exact
agreement with the Matrix string theory result (\ref{eqn:lc_mst}).  In
other words, we have succeeded in reproducing the correct form of the
light-cone gauge type IIA string action in a weakly curved background.

\subsection{Ramond-Ramond couplings}

Turning to the R-R couplings in the action (\ref{eqn:linear_mst}), let
us consider the 3-form couplings, which could potentially give rise to
D2-brane solutions.  If we set the Born-Infeld field to zero, we find
\be
S_{C^{(3)}} = \frac{i}{2\pi g_s}  \int d\tau d\sigma
~\STr \left\{ \frac{\sqrt{\al'}}{R} C^{(3)}_+ + \dot{X}^i C^{(3)}_i \right\},
\label{C^3}
\ee
where we have defined
\bea
C^{(3)}_+ &=& \frac{1}{\sqrt{2}} C^{(3)}_{+ij} [X^j,X^i], \\
C^{(3)}_i &=& \frac{1}{2} C^{(3)}_{ijk} [X^k,X^j] + C^{(3)}_{+-j} [X^j,X^i].
\eea
The couplings given in~\cite{schiappa} differ considerably from the
ones given here, since in~\cite{schiappa} only the $C^{(3)}_{0ij}$ and
$C^{(3)}_{ijk}$ terms were considered. Yet from our analysis it is
clear that the other terms not only contribute, but are
necessary in order to be able to write the $C^{(3)}$ couplings in the
light-cone gauge.

A nice check of the 3-form couplings (\ref{C^3}) is against
the eleven-dimensional supermembrane theory.
The connection between the light-cone treatment of the
eleven-dimensional supermembrane in flat space and Matrix
theory is well-known~\cite{deWit:88}.  One can further argue that
Matrix \emph{string} theory is found by compactifying the light-cone
supermembrane
theory on a circle, the off-diagonal elements of the Matrix string
theory fields being related to the infinite tower of Kaluza-Klein
modes along the compact direction~\cite{sekino:01}.  Given further that the
light-cone supermembrane theory in an arbitrary supergravity background is
known~\cite{BST,deWit:97,deWit:98}, one could in principle derive
Matrix string theory in an arbitrary background using these techniques.  Indeed,
the
Lagrangian density for the light-cone supermembrane in a curved (but
not entirely general) background has been derived in~\cite{deWit:97} and it is
easy to see that the methods of~\cite{sekino:01}
would generate some of the couplings in (\ref{C^3}) of our Matrix
string theory action.  More specifically, the couplings to
$C^{(3)}_{+ij}$ and $C^{(3)}_{ijk}$ are reproduced although, since the
gauge $C^{(3)}_{+-j}=0$ was chosen in~\cite{deWit:97}, the coupling to
these components of $C^{(3)}$ cannot be checked.  Of course, neither can
the couplings to the remaining R-R fields, since it is not known how or,
indeed, if their eleven-dimensional counterparts couple to the supermembrane.

It can be seen from (\ref{eqn:linear_mst}) that as regards the R-R
5-form potential, only the terms of the form $C^{(5)}_{a_1...a_4 9}$
contribute. It is easy to check that the terms involving
$C^{(5)}_{a_1...a_5}$ would couple to the current $I_5$, which was
previously set to zero in order to match the action
(\ref{eqn:D0}) with the action of the multiple D0-brane theory of
\cite{TVR2, myers}. Setting the Born-Infeld vector to
zero, the remaining 5-form R-R field couplings can be written as:
\be
S_{C^{(5)}} = \frac{i}{4\pi g_s} \frac{R}{\sqrt{\al'}} \int d\tau d\sigma
~\STr \left\{ \frac{\sqrt{\al'}}{R} C^{(5)}_{+-} + \dot{X}^i C^{(5)}_{+i}
                  -\dot{X}^i C^{(5)}_{-i} \right\},
\ee
where we have defined
\be
C^{(5)}_\mn = [X^k, X^j] DX^i C^{(5)}_{ijk\mn}.
\ee
Similarly, the $C^{(7)}$ couplings only have contributions involving
terms of the form $C^{(7)}_{a_1 ... a_7}$, since terms of the form
$C^{(7)}_{a_1 ... a_6 9}$ couple to currents for which the explicit
expression is not known, corresponding to $N^{(7)}$ couplings in the
D0-brane action (\ref{eqn:D0}). The $C^{(7)}_{a_1 ... a_7}$ terms
can be written as
\be
S_{C^{(7)}} = \frac{i}{96\pi g_s^3} \frac{R^2}{\al'} \int d\tau d\sigma
~\STr \left\{ \frac{\sqrt{\al'}}{R}\frac{1}{\sqrt{2}} C^{(7)}_{+} +
              \frac{\sqrt{\al'}}{R}\frac{1}{\sqrt{2}} C^{(7)}_{-} +
                \dot{X}^i C^{(7)}_{i} \right\},
\ee
with
\be
C^{(7)}_\mu = [X^n, X^m][X^l, X^k][X^j, X^i] C^{(7)}_{ijklmn\mu}.
\ee

Let us for the sake of completeness also consider the couplings to
$\tb^{(6)}$ and $N^{(7)}$.  As for $C^{(5)}$, only the terms with a
9-component appear in the action. The other components couple to
currents for which the explicit expression is not known.  The
$\tb^{(6)}$ couplings can be written as:
\be
S_{\tb^{(6)}} = \frac{1}{16\pi g_s^2} \frac{R}{\sqrt{\al'}} \int d\tau d\sigma
~\STr \left\{ \frac{\sqrt{\al'}}{R} \tb^{(6)}_{+-} + \dot{X}^i \tb^{(6)}_{+i}
                                                 -\dot{X}^i \tb^{(6)}_{-i}
\right\},
\ee
where
\be
\tb^{(6)}_\mn = [X^l, X^k][X^j,X^i]  \tb^{(6)}_{ijkl\mn},
\ee
and the $N^{(7)}$ couplings are given by
\be
S_{N^{(7)}} = -\frac{1}{16\pi g_s^2} \frac{R^2}{\al'} \int d\tau d\sigma
~\STr \left\{ \frac{\sqrt{\al'}}{R} N^{(7)}_{+-} - \dot{X}^i N^{(7)}_{+i}
                                                 +\dot{X}^i N^{(7)}_{-i}
\right\},
\ee
where
\be
N^{(7)}_\mn = [X^m, X^l][X^k,X^j]DX^i  N^{(7)}_{ijklm\mn}.
\ee

Note that the couplings of the different fields occur at a different order of the
expansion
para\-meter $R/\sqrt{\al'}$.

\sect{T-duality once more: multiple IIB F-strings}

To describe fundamental strings in the type IIB theory, we perform
another T-duality in the $x^9$ direction, as in
(\ref{eqn:t-duality_bgd1}) and (\ref{eqn:t-duality_bgd2}).  As before,
we assume that the worldvolume fields (the currents) do not change, so
the flat action (\ref{eqn:flat_mst}) is unchanged.  The linear action
(\ref{eqn:linear_mst}) becomes
\[
S_{{\rm linear}}^{{\rm IIB}} = \frac{1}{2\pi}\int d\tau d\sigma \frac{\al'}{R^2}
~\STr \left\{
\frac{1}{2} \left( h_\ab -\phi \eta_{ab} \right) I_h^\ab + C^{(2)}_{\a9} I_h^{a9}
+ \frac{1}{2} \left( \phi - h_{99} \right) I_h^{99} - \frac{1}{2}
\left( \phi +  h_{99} \right) I_\phi \right.
\]
\[
-  C^{(2)}_\ab I_s^\ab  - 2h_{\a9}I_s^{\a9}
+ b_{a9} I_0^a - C^{(0)} I_0^9
+ C^{(4)}_{abc9} I_2^{abc} + 3 b_{ab} I_2^{ab9}
+ \frac{1}{12}C^{(4)}_{a_1...a_4} I_4^{a_1...a_4 9}
\]
\be
\left. + \frac{1}{60}\tb_{a_1...a_5 9} I_4^{a_1...a_5}
+ \frac{1}{48}\tb_{a_1\ldots a_6} I_6^{a_1...a_6 9}
- \frac{1}{336} C^{(8)}_{a_1...a_7 9} I_6^{a_1...a_7} \right\},
\label{F1B}
\ee

\noindent which should describe strongly coupled
Matrix strings in the IIB theory, with
the currents as in the appendix, up to the coordinate transformations
(\ref{eqn:rescalings}), and the rescalings $X^i
\longrightarrow \sqrt{\al'} X^i$ and $g_s \longrightarrow
g_s/(2\pi)$.

Note that precisely the same action is obtained if one applies the
S-duality rules (\ref{S-rules1}) and (\ref{S-rules2}) to the D1-brane action
(\ref{eqn:d1_linear}).  Although this might be expected from the
S-duality connection between D- and F-strings in the type IIB theory, it is not
clear \emph{a priori} how such an S-duality should be
done directly, since we are dealing with non-abelian fields; as in
Yang-Mills theory, we cannot rigorously perform a worldvolume duality
transformation to show the S-duality equivalence between the two.

It is worth pointing out however that the S-duality transformation of the
NS-NS 2-form:
\be
\label{bc}
b^{(2)}\longrightarrow -C^{(2)},\qquad \qquad C^{(2)}\longrightarrow b^{(2)},
\ee

\noindent implies that the dynamics will now be
governed by open D-strings, given that the
invariant 2-form field strength that will couple in the worldvolume
is constructed
with $C^{(2)}$ instead
of $b^{(2)}$. Therefore the Born-Infeld field $A$ should transform
into a new worldvolume vector field $A^\prime$ whose abelian component forms
a gauge invariant field strength with the R-R 2-form.
The right S-duality transformation rule should be then

\be
A\longrightarrow -A^\prime, \qquad \qquad A^\prime \longrightarrow A,
\ee

\noindent in order to match (\ref{bc}). This
cannot however be seen explicitly at the level of the linearized
actions that we have considered.

The action above can be rewritten in a more convenient form, filling in the
expressions for the currents as given in the appendix (upon the rescaling).
In particular, for the Chern-Simons action we
have:
\[
S_{{\rm CS}}^{{\rm IIB}} = \frac{1}{2\pi} \int d\tau d\sigma ~\STr \left\{ P
\left[ \frac{\al'}{R^2} b^{(2)} + \frac{g_s\sqrt{\al'}}{R} C^{(0)}\wedge
F + \frac{i\sqrt{\al'}}{g_s R} (\incl_X \incl_X) C^{(4)} + i (\incl_X
\incl_X)b^{(2)} \wedge F \right. \right.
\]
\be
\left. \left. - \frac{1}{2 g_s^2}(\incl_X \incl_X)^2
\tb^{(6)} - \frac{R}{2g_s\sqrt{\al'}}(\incl_X \incl_X)^2 C^{(4)} \wedge F
 + \frac{iR}{6 g_s^3\sqrt{\al'}}(\incl_X \incl_X)^3 C^{(8)}
- \frac{iR^2}{6 g_s^2 \al'}(\incl_X \incl_X)^3 \tb^{(6)}\wedge F
              \vphantom{\frac{i\sqrt{\al'}}{g_s R}}          \right] \right\}.
\label{F1B2}
\ee
Again we  note that only half of the terms necessary to form the pullback
of $C^{(8)}$ are present in the linear action.
Some of the above couplings have been given before in \cite{yolanda}.

\sect{T-duality along a transverse direction: multiple IIA strings
with winding number}

Let us now also describe type IIA fundamental strings with winding
number. The IIA strings that are described by Matrix string theory
carry momentum $p^+$.  This is the charge which is related
by the 9-11 flip to the number of D-particles.  T-duality in the $x^9$
direction gives IIB strings with winding number, which we have
just checked are S-dual to multiple D-strings. Strongly coupled
IIA F-strings with
winding number can then be obtained from IIB strings by performing a
T-duality transformation in a direction transverse to the IIB strings.
These strings with winding number have interesting dielectric
properties that we will discuss in the next section.

Calling $z$ the T-duality direction and $a=(0, i)$,
where now $i= 1, ...,7$,  the linear
action that is
obtained from (\ref{F1B}) is given by:
\[
S_{{\rm linear}}^{{\rm IIA}} = \frac{1}{2 \pi} \int d\tau d\sigma
\frac{\al'}{R^2} ~\STr \left\{
\frac{1}{2} \Bigl(h_{ab} + \eta_{ab} (\phi-\frac{1}{2} h_{zz}) \Bigr) I_h^{ab}
- b_{az} I_h^{az}
- \frac{1}{2} \Bigl( \frac{1}{2} h_{zz} + \phi\Bigr) I_h^{zz} \right.
\]
\[
+ \ C^{(3)}_{a9z}I_h^{az} -\frac{1}{2} \Bigl( h_{99} -\phi +
\frac{1}{2}h_{zz} \Bigr) I_h^{99} - C^{(1)}_9 I_h^{z9}
- \frac{1}{2} \Bigl( \phi - \frac{1}{2} h_{zz} + h_{99}\Bigr) I_\phi -
C^{(3)}_{abz} I_s^{ab} - 2 C^{(1)}_a I_s^{az}
\]
\[
- 2 h_{a9} I_s^{a9} -2 b_{z9} I_s^{z9} + b_{a9} I_0^a
+ h_{z9} I_0^z - C^{(1)}_z I_0^9  + C^{(5)}_{abc9z} I_2^{abc} - 3
C_{ab9} I_2^{abz} + 3 b_{ab} I_2^{ab9} - 6 h_{az} I_2^{az9}
\]
\[
+ \frac{1}{60} N^{(7)}_{a_1 ... a_59z} I_4^{a_1 ... a_5}
+ \frac{1}{12} \tb_{a_1 ... a_4z9} I_4 ^{a_1 ... a_4z}\nn
+ \frac{1}{12} C^{(5)}_{a_1 ... a_4z} I_4^{a_1 ... a_49}
+ \frac{1}{3} C^{(3)}_{abc} I_4^{abcz9}
- \frac{1}{336} C^{(9)}_{a_1 ... a_79z} I_6^{a_1 ... a_7}
\]
\be
\left.
+ \frac{1}{48} C^{(7)}_{a_1 ... a_69} I_6^{a_1 ... a_6z} +
\frac{1}{48}N^{(7)}_{a_1 ... a_6z} I_6^{a_1 ... a_69}
+ \frac{1}{8} \tb_{a_1 ... a_5z} I_6^{a_1 ... a_5z9} \right\} .
\label{F1A*}
\ee
The direction $z$ in which the T-duality is performed appears as an isometry
direction in
the transverse space of the strings. We denote the corresponding Killing vector
as
\be
k^\mu = \delta_z^\mu,
\hsp{2cm}
k_\mu = \eta_{z\mu} + h_{z\mu}.
\ee
In a manner similar to the Kaluza-Klein
monopole, the non-abelian strings do not see this special direction, the
embedding scalar $X^z$ is not a degree of freedom of the strings, but is
transformed under
T-duality into a world volume scalar $\omega$ \cite{EJL}.
This worldvolume scalar forms an invariant field strength with $b_{az}$
(see \cite{yolanda} for the details), and can therefore be associated to
fundamental strings wrapped around the isometry direction $z$, which
themselves end on the Matrix strings.

The action (\ref{F1A*}) can be written in a covariant way as a gauged sigma
model, where gauge covariant derivatives $\cD_\alpha X^\mu$ are used to gauge
away
the  embedding scalar corresponding to the isometry direction \cite{BJO}:
\be
\cD_\alpha X^\mu = D_\alpha X^\mu - k_\rho D_\alpha X^\rho k^\mu,
\ee
with $\alpha = \sigma, \tau$. These gauge covariant derivatives reduce to the
standard
covariant derivatives $D_\alpha X^\mu$ for $\mu \neq z$ and are zero for $\mu =
z$.
The pull-backs that appear in the action of the F-strings with winding are
constructed from these
gauge covariant derivatives. For example,
\bea
\cP \left[ b^{(2)} \right]
              &=& b_\mn \cD X^\mu \cD X^\nu dt dx\\
             & =&  \left( b_{09} + \frac{R}{\sqrt{\al'}} b_{0i} DX^i
                  +\frac{R}{\sqrt{\al'}}  b_{i9} \dX^i
                  +\frac{R^2}{\al'} b_{ij} \dX^i DX^j \right) dt dx. \nonumber
\eea

Filling in the expressions for the currents as given in the appendix (upon the
rescaling)
we can write the Chern-Simons action as:
\[
S_{{\rm linear}} = \frac{1}{2\pi} \int d\tau d\sigma
~\STr \left\{ \cP \left[ -\frac{ g_s\sqrt{\al'}}{R} \incl_k C^{(1)} \wedge  F
+ \frac{\al'}{R^2} b^{(2)}
- \frac{\sqrt{\al'}}{R} k^{(1)} \wedge D\omega + i (\incl_{[X,
\omega]}) k^{(1)} \wedge F \right. \right.
\]
\[
+  \frac{i\sqrt{\al'}}{g_sR} (\incl_{[X, \omega]}) C^{(3)}
+ \frac{i}{g_s} (\incl_X \incl_X)  C^{(3)}\wedge D\omega
+ \frac{R}{g_s\sqrt{\al'}} (\incl_X \incl_X)(\incl_{[X, \omega]})
C^{(3)} \wedge F - \frac{i\sqrt{\al'}}{g_sR} (\incl_X \incl_X)  \incl_k C^{(5)}
\]
\[
+ \frac{R}{2g_s\sqrt{\al'}} (\incl_X \incl_X)^2 \incl_k C^{(5)} \wedge F
+ \frac{1}{g_s^2} (\incl_{[X, \omega]})(\incl_X \incl_X) \incl_k
\tb^{(6)} +  \frac{R}{2 g_s^2\sqrt{\al'}}(\incl_X \incl_X)^2 \incl_k
\tb^{(6)}\wedge D\omega
\]
\[
+ \frac{R^2}{2 g_s\al'} (\incl_{[X, \omega]})(\incl_X \incl_X)^2
          \incl_k \tb^{(6)}\wedge F - \frac{1}{2 g_s^2} (\incl_X \incl_X)^2
\incl_k N^{(7)}
- \frac{iR^2}{6 g_s^2\al'} (\incl_X \incl_X)^3 \incl_k N^{(7)} \wedge F
\]
\be
\left. \left. \vphantom{\frac{ g_s\sqrt{\al'}}{R}} - \frac{iR^2}{2
g_s^3\al'}  (\incl_{[X, \omega]}) (\incl_X \incl_X)^3 C^{(7)}
+ \frac{R}{6 g_s^3\sqrt{\al'}}  (\incl_X \incl_X)^3\incl_k C^{(9)} \right]
\right\},
\label{F1A*CS}
\ee
where we have introduced the following types of interior multiplication:
\bea
&& (i_k \Sigma)_{\mu_1...\mu_p} = k^\rho \Sigma_{\rho\mu_1...\mu_p}
                              =  \Sigma_{z\mu_1...\mu_p},
\nn
&& (\incl_{[X, \omega]} \Sigma)_{\mu_1...\mu_p} = [X^i, \omega]
\Sigma_{i\mu_1...\mu_p}.
\eea
Again only half of the terms to form the pullback of $C^{(7)}$ and  $C^{(9)}$
are present in the linear action (\ref{F1A*}). Some of the couplings
in (\ref{F1A*CS}) have been derived earlier in \cite{yolanda},
applying T- and S-duality relations to the actions of D-branes.
We will see in the next section that
the presence of the isometry direction will enable us to
find dielectric solutions of F-strings expanding into D$p$-branes
with $p$ even.

\sect{Dielectric solutions}

It is well-known in the case of D-branes that some of the linear couplings in
(\ref{eqn:D0}) and (\ref{eqn:d1_linear}) give rise to  stable, dielectric
configurations. In particular it was shown \cite{myers,trivedi:00} that a set of
coinciding
D$p$-branes in the presence of a $(p+4)$-form R-R field strength will
expand into a non-commutative or fuzzy two-sphere, $S_{NC}^2$.  This
configuration is stable and can be identified with a fuzzy
D$(p+2)$-brane of topology $\mathbb{M}^{p+1} \times S_{NC}^2$.  This
dielectric D$(p+2)$-brane has no net (monopole) charge with respect to
the $(p+4)$-form R-R field strength, but it does have a dipole moment.

In this section we will comment on various solutions describing F-strings
expanding into
D-branes.

\subsection{Dielectric D-strings}
It is instructive to review the case of $N$ D-strings expanding into a dielectric
D3-brane in our notation. The action for $N$ coinciding D-strings is given by the
action (\ref{eqn:d1_flat}) for a flat background plus the action
(\ref{eqn:d1_linear})
for the linear couplings, which in the static case ($\dX^i = DX^i = \dA = 0$)
gives rise to the following potential:
\be
V_{{\rm D}1} = \STr \left\{-\frac{1}{4 \lambda^2} [X, X]^2
                      + \frac{i}{3\lambda} X^k X^j X^i F^{(5)}_{09ijk} \right\},
\label{D1pot}
\ee
where, in order to obtain the couplings to the five-form field strength, we
performed a partial integration and a non-abelian Taylor expansion \cite{GM,
myers}
\be
C^{(4)}_{\mnrl}(X) = C^{(4)}_{\mnrl} (0)  + \part_k C^{(4)}_{\mnrl}(0) X^k + ...
\ee
The equation of motion
\be
[ [X^i, X^j], X^j ] + \frac{i \lambda}{2} [X^k, X^j] F^{(5)}_{09ijk} = 0,
\ee
is clearly satisfied for the following choice of $F^{(5)}$ and $X^i$:
\bea
&& F^{(5)}_{09ijk} = f \varepsilon_{ijk},\nn
&& X^i = -\frac{\lambda}{4} f \sigma^i
\hspace{.4cm} \mbox{where} \hspace{.4cm} [\sigma^i, \sigma^j] = 2i \
\varepsilon^{ijk}\sigma^k,
\label{D1-D3}
\eea
where $i,j,k = 1,2,3$.  In other words, three of the transverse coordinates of
the D-string form a $N\x
N$ matrix representation of $SU(2)$. The D-strings have expanded into a fuzzy
two-sphere with radius
\be
R^2 \equiv \frac{1}{N} {\rm Tr} (X^i X^i)  = \frac{\lambda^2}{16} f^2 (N^2 -1),
\label{radius}
\ee
representing a (fuzzy) D3-brane of topology $\mathbb{M}^2 \times S_{NC}^2$, with
no net D3-brane
charge, but with a dipole moment with respect to $F^{(5)}$. The
potential (\ref{D1pot}) for the solution (\ref{D1-D3}) gives
\be
V_{{\rm D}1} = -\frac{\lambda^2}{384} f^4 N(N^2-1),
\label{energy}
\ee
which is clearly negative. The dielectric D3-brane thus has a lower
energy than the original configuration of $N$ coinciding D-strings,
and the latter will spontaneously decay into the former.

\subsection{Dielectric F-strings in type IIB}
It is clear that for the type IIB F-strings, a similar effect will occur. If we
compare the linear couplings (\ref{F1B2}) of the type IIB F-strings
with those of the D-strings (\ref{eqn:d1_pullbacks}), we notice that,
although the former now has a monopole
charge with respect to $b^{(2)}$, the dielectric coupling to $C^{(4)}$ is
identical
to that for the D-strings. Hence the action (\ref{F1B2})
for $N$ coinciding type IIB F-strings gives rise to the same type of potential:
\be
V_{{\rm F}1}^{{\rm IIB}} = \STr \left\{-\frac{1}{4 g_s^2} [X, X]^2
                      + \frac{i}{3g_s} X^k X^j X^i F^{(5)}_{09ijk} \right\},
\label{F1pot}
\ee
where now we performed a non-abelian Taylor expansion
\be
C^{(4)}_{\mnrl}(X) = C^{(4)}_{\mnrl} (0)  + \frac{R}{\sqrt{\al'}}\part_k
C^{(4)}_{\mnrl}(0) X^k + ...
\ee
Clearly, the potential (\ref{F1pot}) has a solution of the same type as
(\ref{D1-D3}):
\bea
&& F^{(5)}_{09ijk} = f \varepsilon_{ijk},\nn
&& X^i = -\frac{g_s}{4} f \sigma^i
\hspace{.4cm} \mbox{where} \hspace{.4cm} [\sigma^i, \sigma^j] = 2i \
\varepsilon^{ijk}\sigma^k.
\label{F1-D3}
\eea
The interpretation now is that the $N$ F-strings have expanded to form a fuzzy
D3-brane with topology $\mathbb{M}^2 \times S_{NC}^2$, the S-dual
of the D1-D3 configuration described above. Again this
is a stable configuration with  radius and energy as in (\ref{radius}) and
(\ref{energy}) respectively.

Similarly, the $(\incl_X \incl_X)^2 \ \tb^{(6)}$ couplings in (\ref{F1B2})
indicate
that F-strings can expand into an NS5-brane with quadrupole moments, this
being the S-dual of a D1-D5 configuration. A solution of $N$ D-strings
expanding into a D5-brane was given
in \cite{trivedi:00}, using the gamma matrices of $SO(5)$. This solution however
turned out to be unstable, and we expect to find a similar unstable solution
here.

\subsection{Matrix string theory: dielectric F-strings with momentum in type IIA}

The case of type IIA Matrix string theory is more subtle.  A solution, claimed to
describe $N$ F-strings expanding into a spherical D2-brane, has been
given in \cite{schiappa}. There, the $C^{(3)}$-couplings were considered in the
static
gauge $\tau = t$, $\sigma = x^9$, rather then in the light cone gauge as in
section~4.
The potential is then
\be
V_{{\rm F}1}^{{\rm IIA}} = \STr \left\{-\frac{1}{4 g_s^2} [X, X]^2
                      + \frac{i}{3g_s} X^k X^j X^i F^{(4)}_{0ijk} \right\},
\ee
which clearly has a minimum of the form (\ref{F1-D3}) with
$F^{(4)}_{0ijk}= f \varepsilon_{ijk}$. However, the question arises as
to whether this
solution really corresponds to a {\it string} expanding into a D2-brane, given
that the
worldvolume of the D2-brane ($(t, x^i)$ in Cartesian coordinates) does not
contain the world sheet
direction, $x^9$, of the string.  How a collection of strings could give rise
to such a solution, where no trace of the spatial component of the worldvolume of
the
strings is found, is unclear.  Moreover, the dipole coupling $(\incl_X
\incl_X) C^{(3)}$ in the Chern-Simons action, that would
describe such a situation, cannot
appear as a pull-back to a two-dimensional worldvolume. Nor is it possible to
construct from the
action (\ref{C^3}) a R-R four-form field strength coupling of the form $F_{09ij}$
or $F_{+-ij}$, which would include the world sheet coordinate of the string in
the dielectric
brane.

Let us take static gauge $X^9 = \sigma$ in the linear action
(\ref{eqn:linear_mst})
so that we can better identify the dielectric couplings that appear
in the Chern-Simons part of this action. In that case, one immediately realizes
that the
couplings cannot be written as a pull-back to a two-dimensional worldvolume.
Moreover,
the different terms can be organized as pull-backs
to a one-dimensional worldvolume which has $X^9$ as an isometric direction.
It turns out that the resulting action describes multiple pp-waves
carrying momentum along the $X^9$ direction, as in \cite{BT}.
Pp-waves carrying momentum in a compactified direction are, of course,
T-dual to fundamental strings wound around this direction
\cite{BEK,bert}. There is a sense, then, in which waves can be
considered as fundamental strings carrying momentum (see
\cite{hull} for a discussion of this point). We have found that
the action that describes type IIA Matrix strings coupled to background
fields reduces to that describing multiple waves when one goes to the
static gauge, which is in agreement with the fact that Matrix strings
carry momentum along the compactified direction.

The interpretation in terms of waves sheds some light on the
dielectric solution for IIA Matrix strings constructed in
\cite{schiappa}.  Recall that the description of multiple waves is in terms
of a gauged sigma model in which the direction of propagation of the waves
is isometric \cite{BT}. Indeed, in this action
one finds a coupling $(\incl_X \incl_X) C^{(3)}$
to the worldvolume \cite{yolanda}, but now the pull-backs include gauge covariant
derivatives
(see the discussion about gauged sigma models in section 6) which gauge
away $X^9$ as a world volume scalar.
The resulting dielectric D2-brane is therefore constrained
to move in the space transverse to the isometric direction.

In summary, we believe that the solution
\bea
&& F^{(4)}_{0ijk} = f \varepsilon_{ijk} ,\nn
&& X^i = -\frac{g_s}{4} f \sigma^i
\hspace{.4cm} \mbox{where} \hspace{.4cm} [\sigma^i, \sigma^j] = 2i \
\varepsilon^{ijk}\sigma^k ,
\label{W-D2}
\eea
presented in \cite{schiappa} corresponds to a set of $N$ gravitational waves
expanding
into a spherical D2-brane, which again is a stable
configuration with radius and energy as in (\ref{radius}) and
(\ref{energy}) \footnote{A similar type of solution, for a magnetic
4-form field strength, has been constructed in \cite{DTV}.}.

We have argued in the introduction that we do not expect to find
stable solutions describing F-strings expanding into a cylindrical D2-brane.
Since we
cannot find a coupling of the F-strings to either $F_{09ij}$ or
$F_{+-ij}$, which would give rise to such a dielectric brane, this
expectation is confirmed.  Such a configuration simply
does not exist in the type IIA Matrix string theory.  Although we find
a possible cylindrical D2-brane solution below, it is unstable.
Whether it corresponds to the unstable abelian D2-brane solution is
unclear, however.

\subsection{Dielectric F-strings with winding in type IIA}

Coinciding F-strings with winding can expand into a
spherical D4-brane in
the presence of an R-R 6-form field strength. We can see this by analysing the
$C^{(5)}$ couplings in the action (\ref{F1A*}), which can be rewritten as the
following coupling to the R-R six-form field strength:
\be
S_{{\rm F}1}^{{\rm IIA}}  \sim \frac{i}{3 g_s}  \int d\tau d\sigma \  {\rm STr}
\Bigl\{
X^k X^j X^i F^{(6)}_{09zijk} \Bigr\}\ .
\ee
It is then clear that a stable solution of the potential
\be
V_{{\rm F}1}^{{\rm IIA}} = \STr \left\{-\frac{1}{4 g_s^2} [X, X]^2
                      + \frac{i}{3 g_s} X^k X^j X^i F^{(6)}_{09zijk}
\right\},
\label{F1potA}
\ee
is given by
\bea
&& F^{(6)}_{09zijk} = f \varepsilon_{ijk} ,\nn
&& X^i = -\frac{g_s}{4} f \sigma^i
\hspace{.4cm} \mbox{where} \hspace{.4cm} [\sigma^i, \sigma^j] = 2i \
\varepsilon^{ijk}\sigma^k ,
\label{F1-D4}
\eea
and corresponds to $N$ F-strings expanding into a D4-brane, which is
itself wrapped around the isometry direction $z$. We believe this
solution corresponds to the (smeared) supergravity
solution of an F-string expanding into a D4-brane, the T-dual of the fully
localized
F1-D6 solution given in \cite{emparan:01, brecher:01}.

Coinciding F-strings with winding also have couplings to the R-R
3-form field potential,
which seems to suggest an expansion into a cylindrical D2-brane. Yet these
$C^{(3)}$-couplings have an unusual form, due to the presence of the world volume
scalar field $\omega$. In the static case the potential is given by
\be
V_{{\rm F}1}^{{\rm IIA}} = \STr \left\{-\frac{1}{4 g_s^2} [X, X]^2
-\frac{1}{2 g_s^2}[X,\omega]^2-\frac{i}{6 g_s}
\Bigl(
[X^i,X^j]\omega+[X^j,\omega]X^i+[\omega,X^i]X^j\Bigr)F^{(4)}_{09ij}\right\}.
\label{F1Apot}
\ee
It is easy to check that the following Ansatz
\bea
&& F^{(4)}_{09ij} = f \varepsilon_{ij} ,\nn
&& [X^i,X^j]=0, \qquad [X^i,\omega]=i g_s f \varepsilon^{ij}X^j
\label{F1A-D2}
\eea
where $i,j=1,2$, is a solution of the equations of motion
\bea
&& {[} [X^i, X^j], X^j {]}+{[}[X^i,\omega],\omega{]}-i g_s
     [X^j, \omega] F^{(4)}_{09ij} = 0 , \nn
&& [[X^i,\omega], X^i]-\frac{i}{2} g_s
     [X^i,X^j ] F^{(4)}_{09ij} = 0 .
\eea
The solution (\ref{F1A-D2}) is the algebra of the symmetry group of a cylinder.
A similar
solution was recently given in \cite{Bena} using D0-brane matrix degrees
of freedom\footnote{This can be done by adding a small number of D0-branes
to the F1-D2-brane system and taking then the D0-brane
density to zero.}.  This brane has no net monopole
charge with respect to the $F^{(4)}$ field strength but carries
a dipole moment in the XZ and YZ planes.

{}From (\ref{F1A-D2}), the worldvolume of the D2-brane is $(t,x^9,x^i)$ in
Cartesian coordinates, which suggests a cylindrical worldvolume.
However, the proper interpretation is not entirely clear since the
solution explicitly involves the worldvolume scalar field $\omega$.
It is possible that this solution is dual to the unstable solution of
the abelian D2-brane theory, since it turns out to correspond to a maximum
of the potential and is therefore unstable.  The only stable radius for the
cylinder is
\be
R^2 = \frac{1}{N} {\rm Tr} (X^i X^i) = 0,
\ee
as might be expected from the analysis of \cite{emparan:01,
brecher:01}.  There is another remark we would like to make: an
explicit representation of the cylindrical algebra (\ref{F1A-D2}) is~\cite{Bena}
\be
\omega_{ab}= g_s f \delta_{ab}b,\qquad X^1=R(a + a^\dagger), \qquad
X^2=iR (a^\dagger-a)
\ee
\be
a_{ab}=e^{-i\Omega t}\delta_{a-1,b},\qquad a^\dagger_{ab}=e^{i\Omega t}
\delta_{a+1,b},
\ee
where $a,b=1,\ldots ,N$ and for some arbitrary phase $\Omega$, is only
possible for the infinite-dimensional case.  There are no
finite-dimensional representations of (\ref{F1A-D2}), which
makes the solution rather unphysical, requiring
an infinite number of F-strings for the effect to take place.

\sect{Discussion}

The dielectric effect of Myers~\cite{myers} is not limited to
D-branes, but should be observable in a theory of fundamental strings
too.  To describe such dielectric F-strings, we are led to a
consideration of Matrix string theory, which captures type IIA superstring theory
in the light-cone gauge, together with extra degrees of freedom
representing D-brane states.

We have shown explicitly that the Matrix theory action in a weakly
curved background reproduces the lowest order expansion of the
combined non-abelian Born-Infeld-Chern-Simons of Taylor and van
Raamsdonk~\cite{TVR2} and Myers~\cite{myers}.  Using T-duality and the 9-11 flip,
we have derived the Matrix string theory action in a weakly curved
background and shown that, in the weak string coupling limit, the result
reproduces light-cone
gauge string theory.  We have not been able to find explicit
solutions describing dielectric Matrix strings expanding into a D2-brane in
this theory, and have argued that this is to be expected.  We have
further argued that the solution found in~\cite{schiappa} (in which the static,
as opposed to
the light-cone, gauge was considered) in some sense describes
dielectric particles or pp-waves, carrying momentum.

We have derived a, strongly coupled, type IIB Matrix string theory, 
S-dual to the
D-string theory, and a, strongly coupled, type IIA theory of Matrix 
strings with winding.
As regards the former, the standard D1-D3 dielectric solution of the
D-string theory equally well describes a collection of F-strings
expanding into a D3-brane and other, albeit unstable, solutions of
the D-string theory carry through in a similar way.  The theory of Matrix
strings with winding exhibits an isometry direction, in a manner
similar to the theory of the Kaluza-Klein monopole, and one can find
dielectric D4-brane solutions which are smeared over this direction.
Finally, we have mentioned a cylindrical solution, the interpretation
of which is somewhat unclear since it involves the worldvolume scalar
$\omega$.

On a different note, one might expect to find solutions of the type IIA Matrix
string
theory corresponding to the cylindrical supertubes of~\cite{MT,EMT}.  As far
as both the D2-brane theory and the supergravity solutions are
concerned, these involve a cylindrical D2-brane with dissolved
F-string and D0-brane charge.  The configuration is supported from
collapse by an angular momentum in the circle direction.  It should be
clear that possible solutions of Matrix string theory corresponding to
such supertubes would then involve
couplings to both $C^{(3)}$ and $C^{(1)}$.  As far as the
D0-branes are concerned, however, the D2-brane worldvolume is a non-commutative
cylinder~\cite{BL, BL2}.  Then, since the F-strings lie in the direction
along the cylinder, the worldsheet direction would also have to be
non-commutative --- and this cannot occur in Matrix string theory.

The proper determination of both the Matrix and the Matrix string
theories in an arbitrary background remains an open problem.  Of
course, they could be determined by duality transformations of
multiple D-brane theories in an arbitrary background, but the correct form of
the latter is also only poorly understood; and so are the duality
transformations for these non-abelian actions.  We are tempted to think that
one might make progress via a comprehensive treatment of light-cone
gauge supermembrane theory in a general background.

\vspace{1cm}
\noindent
{\bf Acknowledgements}\\
We wish to thank Roberto Emparan for collaborations during the
initial stages of this project.  DB and BJ would further like to thank
Robbert Dijkgraaf, Clifford Johnson, Paul Saffin and Douglas Smith for useful
discussions.  DB is supported in part by the EPSRC grant
GR/N34840/01.

\appendix

\renewcommand{\theequation}{\Alph{section}.\arabic{equation}}

\sect{Appendix: D-string currents}

T-duality applied to the D0-brane currents is explained in the text.
We give the final results here for completeness.  With $i,j=1,\ldots,8$, we have
\bea
I_0^0 &=& \unity, \\
I_0^i &=& \dot{X}^i, \\
I_0^9 &=& \lambda \dot{A}, \\
I_2^{0ij} &=& -\frac{i}{6\beta} [X^i, X^j], \\
I_2^{09i} &=& -\frac{\la}{6\beta} DX^i, \\
I_2^{ijk} &=& -\frac{i}{6\beta} \left( \dot{X}^i [X^j,X^k]
+ \dot{X}^j [X^k,X^i] + \dot{X}^k [X^i,X^j] \right), \\
I_2^{9ij} &=& -\frac{\la}{6\beta} \left( i \dot{A} [X^i, X^j] - \dot{X}^i DX^j
+ \dot{X}^j DX^i \right), \\
I_4^{0ijkl} &=& -\frac{1}{2\beta^2} \left
( [X^i,X^j][X^k,X^l] + [X^i,X^k][X^l,X^j] + [X^i,X^l][X^j,X^k]
\right), \\
I_4^{09ijk} &=& \frac{i\la}{2\beta^2} \left( DX^i [X^j,X^k] + DX^j
[X^k,X^i] + DX^k [X^i,X^j] \right), \\
I_4^{ijklm} &=& -\frac{15}{2\beta^2} \dot{X}^{[i} [X^j,
X^k][X^l,X^{m]} ], \\
I_4^{9ijkl} &=& -\frac{3\la}{2\beta^2}  \left( \dot{A}
[X^{[i},X^j][X^k,X^{l]}] + 4 i \dot{X}^{[i} [X^j, X^k] DX^{l]}
\right), \\
I_6^{0ijklmn} &=& \frac{i}{\beta^3} [X^{[i},X^j][X^k,X^l][X^m,X^{n]}], \\
I_6^{09ijklm} &=& \frac{\la}{\beta^3} DX^{[i} [X^j, X^k][X^l,X^{m]}], \\
I_6^{ijklmnp} &=& \frac{7i}{\beta^3} \dot{X}^{[i}
[X^j,X^k][X^l,X^m][X^n,X^{p]}], \\
I_6^{9ijklmn} &=& \frac{\la}{\beta^3} \left( i \dot{A}
[X^{[i},X^j][X^k,X^l][X^m,X^{n]}] - 6 \dot{X}^{[i} [X^j, X^k][X^l,X^m]
DX^{n]} \right).
\eea
In terms of pullbacks of background fields, we have (for $\beta=\la$ and
$a,b, =0, ..., 8$)
\be
C^{(0)} I_0^9 ~ dtdx = \la P[ C^{(0)} ] \wedge F,
\ee
\be
\left( C^{(2)}_{a9} I_0^a + 3 C^{(2)}_{ab} I_2^{ab9} \right) dtdx = P
[ C^{(2)} ] +\frac{i}{\la} P[ \left(\incl_X
\incl_X\right) C^{(2)} ] \wedge F,
\ee
\be
\left( C^{(4)}_{abc9} I_2^{abc} + \frac{1}{12} C^{(4)}_{a_1 \ldots a_4}
I_4^{a_1 \ldots a_4 9} \right) dtdx = \frac{i}{\la} P[ \left
( \incl_X \incl_X \right) C^{(4)} ] - \frac{1}{2\la} P[ \left
( \incl_X \incl_X \right)^2 C^{(4)} ] \wedge F,
\ee
\be
\left( \frac{1}{60} C^{(6)}_{a_1 \ldots a_5 9} I_4^{a_1 \ldots a_5}
+ \frac{1}{48} C^{(6)}_{a_1 \ldots a_6} I_6^{a_1 \ldots a_6 9} \right)
dtdx = - \frac{i}{2\la^2} P[ \left( \incl_X \incl_X \right)^2
C^{(6)} ] - \frac{i}{6\la^2} P[ \left( \incl_X \incl_X \right)^3
C^{(6)} ] \wedge F,
\ee
\be
\left( \frac{1}{336} C^{(8)}_{a_1 \ldots a_7 9} I_6^{a_1 \ldots a_7} +
\ldots \right) dtdx = -\frac{i}{6\la^3} P[ \left( \incl_X \incl_X \right)^3
C^{(8)} ] + \frac{1}{24\la^3} P[ \left( \incl_X \incl_X \right)^4
C^{(8)} ] \wedge F,
\ee
giving the action (\ref{eqn:d1_pullbacks}) in the text.  The dots
stand for the unknown
coupling to $C^{(9)}$ which, of course, could be determined by
demanding that it can be rewritten in terms of the pullback.  The
dilaton current is
\[
I_{\p} = \unity - \frac{1}{2} \dot{X}^2 \left( \unity + \quarter
\dot{X}^2 + \frac{\la^2}{4\beta^2} DX^2 + \frac{\la^2}{4} \dot{A}^2 -
\frac{1}{8\beta^2}
[X,X]^2  \right) - \frac{1}{8\beta^4} [X,X]^4 -
\frac{\la^2}{2\beta^2}\left( \dot{X} \cdot DX \right)^2
\]
\[
+ \frac{\la^2}{2\beta^2} DX^2 \left( \unity - \quarter \dot{X}^2 - \frac{\la^2}{4\beta^2}
DX^2 +
\frac{\la^2}{4} \dot{A}^2 - \frac{1}{8\beta^2} [X,X]^2  \right) -
\frac{\la^2}{2\beta^4} DX^i [X^i,X^j][X^j,X^k]DX^k
\]
\[
- \frac{\la^2}{2} \dot{A}^2 \left( \unity + \quarter \dot{X}^2 -
\frac{\la^2}{4\beta^2} DX^2 +
\frac{\la^2}{4} \dot{A}^2 - \frac{1}{8\beta^2} [X,X]^2
\right) + \frac{1}{2\beta^2} \dot{X}^i [X^i,X^j][X^j,X^k]\dot{X}^k
\]
\be
- \frac{1}{4\beta^2} [X,X]^2 \left( \unity - \quarter \dot{X}^2 +
\frac{\la^2}{4\beta^2} DX^2 -
\frac{\la^2}{4} \dot{A}^2 - \frac{1}{8\beta^2} [X,X]^2  \right) -
\frac{i\la^2}{\beta^2} \dot{A}DX^i[X^i,X^j]\dot{X}^j,
\ee
and the remaining NS-NS currents are
\bea
I_h^{00} &=& \unity + \frac{1}{2} \dot{X}^2 + \frac{\la^2}{2\beta^2} DX^2 +
\frac{\la^2}{2} \dot{A}^2 - \frac{1}{4\beta^2} [X,X]^2, \\
I_h^{0i} &=& \dot{X}^i \left( \unity + \frac{1}{2} \dot{X}^2 +
\frac{\la}{2} \dot{A}^2 + \frac{\la^2}{2\beta^2} DX^2 - \frac{1}{4\beta^2}
[X,X]^2
\right) \nn
&& \qquad \qquad \qquad \qquad \qquad \qquad  -\frac{i}{\beta^2} [X^i,X^j] \left(
\la^2 \dot{A} DX^j - i [X^j,X^k]\dot{X}^k \right), \\
I_h^{09} &=& \la \dot{A} \left( \unity + \frac{1}{2} \dot{X}^2 + \frac{\la^2}{2}
\dot{A}^2 - \frac{\la^2}{2\beta^2} DX^2 - \frac{1}{4\beta^2} [X,X]^2
\right) + \frac{i\la}{\beta^2} DX^i [X^i,X^j] \dot{X}^j, \\
I_h^{ij} &=& \dot{X}^i \dot{X}^j - \frac{\la^2}{\beta^2} DX^i DX^j -
\frac{1}{\beta^2}
[X^i,X^k][X^k,X^j], \\
I_h^{9i} &=& \la \dot{A} \dot{X}^i - \frac{i\la}{\beta^2} [X^i,X^j] DX^j, \\
I_h^{99} &=& \la^2 \dot{A}^2 - \frac{\la^2}{\beta^2} DX^2, \\
I_s^{0i} &=&  -\frac{1}{2} \left( -\frac{\la^2}{\beta} \dot{A} DX^i +
\frac{i}{\beta}
[X^i,X^j] \dot{X}^j \right), \\
I_s^{09} &=& -\frac{\la}{2\beta} \dot{X} \cdot DX, \\
I_s^{ij} &=& \frac{i}{2\beta} [X^i,X^j] \left(\unity -
\frac{1}{2} \dot{X}^2 - \frac{\la^2}{2} \dot{A}^2 + \frac{\la^2}{2\beta^2} DX^2 -
\frac{1}{4\la^2} [X,X]^2 \right)
- \frac{1}{\beta} \dot{X}^{[i} [X^{j]},X^k] \dot{X}^k  \nn
&& + \frac{\la^2}{\beta} \dot{A} \dot{X}^{[i}
DX^{j]} + \frac{i\la^2}{\beta^3} DX^{[i} [X^{j]},X^k] DX^k -
\frac{i}{2\beta^3} [X^i,X^k][X^k,X^l][X^l,X^j], \\
I_s^{9i} &=& \frac{\la}{2\beta} DX^i \left(\unity -
\frac{1}{2} \dot{X}^2 + \frac{\la^2}{2} \dot{A}^2 - \frac{\la^2}{2\beta^2} DX^2 -
\frac{1}{4\beta^2} [X,X]^2 \right) \nn
&& \qquad \qquad - \frac{i\la}{2\beta} \dot{A} [X^i,X^j] \dot{X}^j +
\frac{\la}{2\beta} \dot{X}^i
DX \cdot \dot{X} - \frac{\la}{2\beta^3} [X^i,X^j][X^j,X^k]DX^k.
\eea


\end{document}